# Chapter III. The Light of Existence[1]

Brian Whitworth
E-mail: bwhitworth@acm.org

*"There is a theory which states that if anyone discovers exactly what the Universe is for and why it is here, it will instantly disappear and be replaced by something even more bizarre and inexplicable. There is another theory which states that this has already happened."* (Adams, 1995)

## ABSTRACT

This chapter derives the properties of light from the properties of processing, including its ability to be both a wave and a particle, to respond to objects it doesn't physically touch, to take all paths to a destination, to choose a route after it arrives, and to spin both ways at once as it moves. Here a photon is an entity program spreading as a processing wave of instances. It becomes a "particle" if any part of it overloads the grid network that runs it, causing the photon program to reboot and restart at a new node. The "collapse of the wave function" is how quantum processing creates what we call a physical photon. This informational approach gives insights into issues like the law of least action, entanglement, superposition, counterfactuals, the holographic principle and the measurement problem. The conceptual cost is that physical reality is a quantum processing output, i.e. virtual.

## INTRODUCTION

This chapter proposes that light, or more exactly electro-magnetism, is the first physical existence, half-way between the nothing of space and the something of matter. Space in the last chapter was a null program, light in this chapter is space on the move, and matter in the next chapter is light in an endless reboot. Here, rather than the physical world being real and the quantum world imaginary, the physical world is a virtual reality created by quantum processing. This may dismay some, but the physical world as a processing output is not illogical, unscientific or untestable. It isn't illogical because seeing the world as physical no more proves it is so than seeing the sun going round the earth proves that is so. It isn't unscientific because it is a statement about the physical world subject to empirical investigation. It isn't untestable because we can compare how the world behaves with how information behaves.

This theory is nothing less than that all physics, including time, space, mass, charge, energy, spin and gravity, is the output of a processing network Wilczek calls *The Grid,* which is the:

"... *primary world-stuff*" (Wilczek, 2008) p74.

This grid is not what we see but what outputs what we see. Quantum theory and relativity describe its operations, as quantum pixels are set at the finite rate we call the speed of light, and its architecture involves what some call the "atoms of space" (Bojowald, 2008). Yet our processing is just a feeble analogy, as even to simulate the behavior of a few hundred atoms a conventional computer:

"... *would need more memory space that there are atoms in the universe as a whole, and would take more time to complete the task than the current age of the universe.*" (Lloyd, 2006) p53.

Only quantum computers approach this power, and do so by tapping the same grid source. The grid proposed is not the computer hardware we know. It creates all hardware. It is the *original existence* that created the physical universe as a local reality[2] at the big bang, and maintains it to this day. If *this* grid stopped processing the physical universe would disappear like an image on a screen turned off.

In the previous chapter, space was the three-dimensional inner surface of a hyper-bubble[3] that has been expanding since the big bang. Light is now an information wave at right angles to that surface, i.e. orthogonal to three dimensional space (Whitworth, 2010). With this concept and the general properties of information, this chapter considers why light:

1. *Never slows or weakens.* Why doesn't light fade, even after billions of years?
2. *Has a constant speed.* Why is the speed of light a constant?
3. *Comes in packets.* Why must light come in minimum energy quanta?
4. *Moves like a wave but arrives as a particle.* How can light be both a wave and a particle?
5. *Always takes the fastest path.* How can photons know *in advance* the fastest route?
6. *Chooses a path after it arrives.* Is this backwards causation?
7. *Can "detect" objects it never physically touches.* How can non-physical knowing occur?
8. *Entirely passes a filter at a polarization angle*? How does *all* the photon get through?
9. *Spins on many axes, and in both ways, at once.* How do photons "spin"?

These, and other unexplained properties of light, fall naturally out of the model proposed.

## LIGHT AS PROCESSING

### Background

In the seventeenth century, Huygens saw light beams at right angles go right through each other and concluded they must be waves, as if they were objects like arrows, they would collide. He saw light as an expanding wave front, where each strike point was the centre of a new little wavelet, traveling outwards in all directions. As the wavelets spread, he argued, they interfere, as the trough of one wave cancels the crest of another. The end result is a forward moving envelope that at a distance from the source acts like a "ray" of light (Figure 1a). Huygen's principle, that each *wave front point is a new wavelet source expanding in all directions,* explained reflection, refraction and diffraction. Newton's idea of bullet-like corpuscles traveling in straight lines explained only reflection and refraction (Figure 1b), yet his simpler idea carried the day.

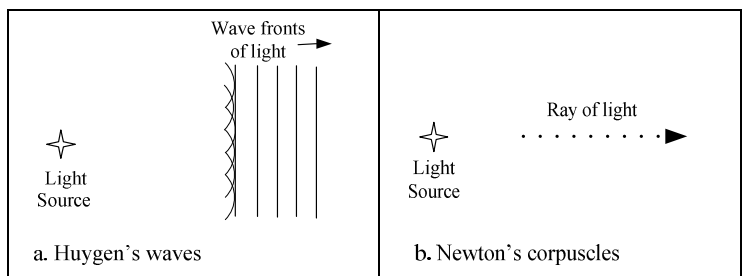

*Figure 1. a. Huygen's wave front vs. b. Newton's corpuscles*

Two hundred years later Maxwell again argued again that light is a wave, as it has a frequency and wavelength, but then Planck and Einstein argued equally convincingly that it comes in particle-like packets. The theory of light has swung from Huygen's waves, to Newton's corpuscles, to Maxwell's waves, to Planck packets. Today, physics *pretends* that light is both wave and particle, even though that is impossible. Three centuries after Newton, the question "*What is light?*" is as controversial as ever. As Einstein commented to a friend just before he died:

---

[2] A local reality, like a simulated world, appears real to its inhabitants but is contained by another reality which generates it. In contrast, an objective reality exists in and of itself and is not contained by anything.

[3] As a circle rotates to give a sphere, so a sphere rotated is a hyper-sphere. While hard to imagine, it is mathematically true that a four-dimensional hyper-sphere has a three-dimensional inner surface.

"*All these fifty years of conscious brooding have brought me no nearer to the answer to the question 'What are light quanta?' Nowadays every Tom, Dick and Harry thinks he knows it, but he is mistaken.*" (Walker, 2000) p89

The point still applies today, i.e. we *really* don't know what light is made of.

## What is light?

In current physics, light vibrates an electro-magnetic field that fills all space, alternately setting positive and negative electric and magnetic potentials at right angles[4]. This wave oscillating slowly is radio and television, faster is heat and visible light, and very fast is x-rays and nuclear rays (Figure 2).

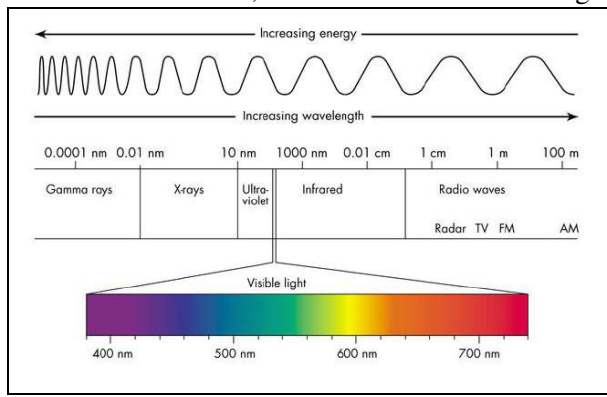

The visible light we see is the part of the spectrum that vibrates at about a million-billion times a second. Low frequency radio waves vibrate a few times a second, while gamma rays oscillate a billion times faster than visible light. From now on, for simplicity, the term light will reference any electro-magnetic vibration. Modern lasers can produce one pulse of light at one frequency in one polarization plane, i.e. a *photon*.

A ray of light can have many photons, each polarized on a different plane through its axis of movement. We know light is a wave because separately visible but out-of-phase photons can interfere to give absolute darkness. A flashlight beam can't do this, but laser generate polarized photons that are individually visible can combine to give absolute darkness. This *light + light = darkness* is only possible for waves.

*Figure 2. The electro-magnetic spectrum*
(http://www.antonine-education.co.uk/)

The amplitude of light is a sine wave, expressed mathematically as a rotation in imaginary space, outside 3D real space. The same mathematics describes a wave on a 2D pond surface. A wave is an oscillation between the force of gravity and water elasticity - one force pushes the water up and an opposing force accelerates it back down again.

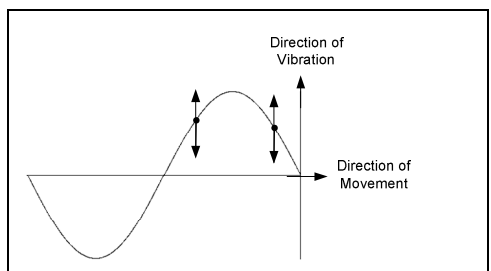

First the surface is flat, then a water molecule is pushed say up, then gravity pulls it back down, then water elasticity pushes it back up, etc. (Figure 3). Viewed from the side, the wave *really* just moves water molecules up and down, as a cork just bobs as a wave passes. The wave that moves across the surface is an interaction pattern, but no physical water moves in the wave direction, until of course it hits a shore.

*Figure 3. Wave particles accelerate up and down on the movement surface*

## What does light vibrate?

If light is a wave and if waves vibrate a medium, then light must have a medium. *Something must move to create light.* Yet there is no physical ether, so physics must simply declare:

"*… we accept as nonexistent the medium that moves when waves of quantum mechanics propagate.*" (Laughlin, 2005) p56.

Light is a wave oscillating between opposing electric and magnetic fields, but no substance is given to these frictionless fields. Electric field changes are proposed to cause the magnetic field changes that cause the electric changes and so on, in a "*… self-renewing field disturbance.*" (Wilczek, 2008)

---

[4] The next chapter takes electric and magnetic fields as aspects of one thing, as is also generally believed.

p212, begging the question of what renews the fields that renew? That an electric force powers a magnetic force which powers the electric force is like Peter paying Paul's bill and Paul paying Peter's bill. With such logic, I could borrow a million dollars and pay nothing back. That light moves at the fastest possible speed but never tires is like a man living extravagantly with no visible means of support. Ancient light, like cosmic background radiation, traveled the universe for billions of years to our telescopes but still arrives at the speed of light. It is colder, as space expanding expanded its wavelength, but its amplitude is undiminished, so a quantum of old light arrives as "fresh" as when it left in energy terms. If light is physical, it has discovered the secret of perpetual motion.

Electro-magnetic waves vibrate forever but physical waves, by the second law of thermodynamics, lose energy by friction with no exceptions[5]. Any wave that vibrates a physical medium must eventually fade, by the inevitable friction of moving physical matter up and down. So the idea of light as a *frictionless wave of nothing* is like no physical wave we know, but that invisible fields in empty space mediate waves that vibrate forever works brilliantly. The hard part is to believe that vibrating nothing (space) can create something (light).

Now suppose a photon program is transmitted by a non-physical processing network. If the grid is idle we see empty space and if it runs this program we see light. Light as a processing wave can then, by the nature of information, be frictionless. Water waves fade, as they move physical water molecules, but a network program uses processing that always runs anyway[6]. An "idle" computer still runs, whether its processing is used or not. Likewise, a processing grid that outputs physicality must be always active, whether the output is a photon (something) or space (nothing). A photon is then everlasting because it is a program sustained by ongoing grid processing[7].

### The speed of space

Einstein deduced the speed of light from how the world behaves, not from how it works:

"… *the speed of light is a constant because it just is, and because light is not made of anything simpler*." (Laughlin, 2005) p15

Why the speed of light in our universe *must be* constant is not explained[8]. Here it is the grid cycle rate, so nodes pass photon programs on at that rate. In a vacuum, it is the idle rate, with nothing else to do. In transparent matter like glass, the grid is also processing the matter, so it will pass on more slowly, as a computer slows down if other programs are running. We *say* the medium of light is glass, but really it is the grid, which also mediates the glass.

The grid cycle rate keeps photons in lock-step sequence behind each other, like the baggage cars of a train driven by the same engine. If this engine slows down under load, as when near a massive object, photons go slower *but still keep the same order,* so no photon can ever overtake another. If it were not so, one could see an object leave, then see it arrive! Temporal causality depends critically upon photons keeping in sequence, which the grid processing engine rigorously maintains.

How the effortless transmission of light contrasts with the forced movement of matter is discussed in Chapter 5. Inherently stationary matter can have a movement property as it needs energy to *start* its motion, but light needs energy to *stop* its automatic grid transmission. The speed of light isn't a property of light at all but of the grid that transmits it. It should be called the speed of space.

---

[5] Planets orbit forever, but the gravity causing this is here derives from the same grid source as light.

[6] Processing, by its nature, must continually run, e.g. an "idle" computer still runs a null cycle, so it is not really doing nothing. Likewise, empty space is here never really empty.

[7] The grid doesn't exist in our (virtual) space - its architecture defines space and its cycles define time.

[8] To say a photon has no mass so goes at the speed of light doesn't explain why there is a maximum speed in the first place. Why can't a photon go at the speed of light plus one? In contrast, a virtual world run by a processor must have a cycle rate, e.g. a 5GHz computer runs 5,000,000,000 cycles per second. Now the speed of light is not just an arbitrary discovery of our universe but an inevitable consequence of its ongoing creation.

### Vibrating *on* space

Does light oscillate in a physical direction, as sound does? To an objective realist the question seems senseless, as how else could it vibrate? Yet consider the facts. Sound is a longitudinal wave that expands and contracts physical air molecules in its travel direction, so there is no sound in empty space. Light is a transverse wave, oscillating at right angles to its line of travel, that still shines in space, or we couldn't see the stars at night. So how can it vibrate anything physical? Also, if light vibrates transversely, it can't have a physical direction because space is *isotropic,* i.e. it has no absolute directions, as "up" from one view is "down" from another. So light vibrating transversely in space *cannot possibly* give the absolute positive and negative values of electro-magnetism.

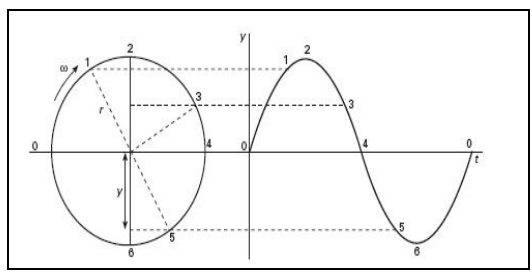

*Figure 4. A turning and moving circle maps to a sine wave*

Charge can only be absolute if light vibrates *outside space,* i.e. in no spatial direction. Space as a 3D surface in a 4D hyper-space allows absolute dimples and dents, like on the surface of a 3D ball. If these displacements form a wave, it can no more leave the surface of space than a water wave could leave a lake surface. Light as a hyper-surface oscillation would move in three spatial directions but vibrate in a dimension imaginary to us.

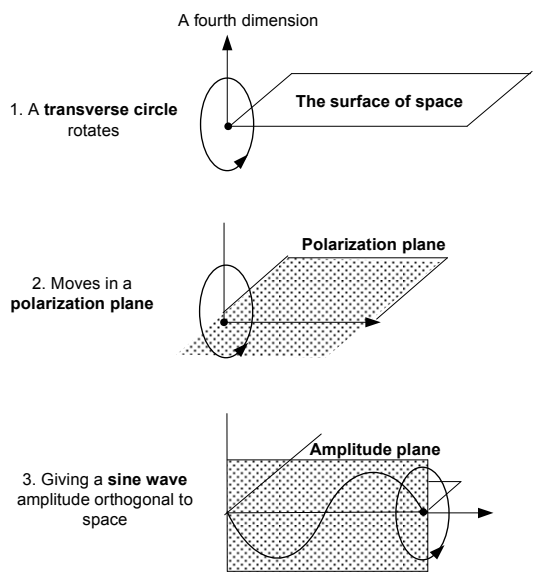

*Figure 5. A transverse circle in quantum space*

Its amplitude would vary as a sine wave, which complex mathematical theory maps to a point rotating into a dimension imaginary to us (Figure 4). Consider Abbot's story of Flatlanders, living on a flat surface (Edwin Abbott, 1884). A ball passing through their world would look to them like a set of circles expanding and contracting, so they might postulate an extra dimension to explain it.

If a turning transverse wheel moved across their world surface, its point amplitude would create a sine wave. If we are 3D Flatlanders in a 4D quantum space, the imaginary dimension of complex number theory would actually exist. A rotating transverse circle moving across a 3D surface would create a sine wave with respect to a polarization plane (Figure 5). In this model, the transverse circle is informational, i.e. a set of values with no inherent nature except relative to other values. These values are proposed to be our physical existence in this world. We can't move into the imaginary direction if it defines our existence. A photon as a transverse rotation moving on the surface of space is a three-dimensional structure that projects two dimensions into our space (its polarization plane).

So complex numbers describe electro-magnetism well because light really does rotate into a dimension outside our space (Figure 6). The core operation of complex numbers, a rotation outside space, then actually occurs[9]:

"*In quantum mechanics there **really are** complex numbers, and the wave function **really is** a complex-valued function of space-time.*" (Lederman & Hill, 2004) p346

In other words, the mathematics of complex number theory represents fact not fiction.

### Fields and dimensions

According to Feynman:

"*A real field is a mathematical function we use for avoiding the idea of action at a distance.*" (Feynman, Leighton, & Sands, 1977) Vol. II, p15-7

We say the earth holds its moon in orbit by the distant acting force of its gravitational field. Such a field permeates all space, to give a value to each point in it, i.e. *adds a degree of freedom to space.* That an electric field has a value even if no charges are present implies that something beyond space exists. Postulating other dimensions, as physics also does, also adds values to all points of space, but while adding a new field is easy, new dimensions compound, e.g. string theory's ten extra dimensions give an estimated $10^{500}$ possible architectures. Fields, as mathematical fictions can multiply regardless of explanatory cost, but dimensions interact, as the problems of string theory show (Woit, 2007).

Fields are now so accepted that we forget they are explanatory constructs, not observed reality. No-one has ever seen gravity - we only see its effects. Modern field theory pretends they are physical by invoking virtual "particles" to cause their effects. Electro-magnetic fields are said to operate by virtual photons, weak nuclear fields by W and Z bosons, strong nuclear fields by gluons, gravity by gravitons and the Higgs field by the Higgs "God" particle. This reassures us that only particles can cause forces, though they are observed only as brief spectral events, i.e. they aren't particles in any normal sense.

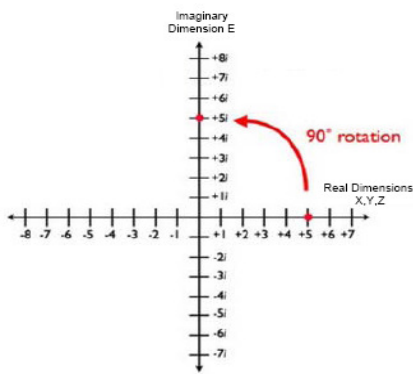

*Figure 6. Complex rotations*

Here, particles are events not things, programs running on a network not billiard balls on a pool table. A field as a useful theoretic device can be replaced by another that works as well, i.e. fields can be modeled as extra dimensions, as string theory does. If each new field implies a new dimension, the grand physics goal of field unification equates to describing gravity, electricity, magnetism, and strong and weak fields using only a single extra dimension. In this model, that dimension is the unseen and unseeable one into which light, or electro-magnetism, vibrates. Mathematically, it is just a degree of freedom beyond space, into which values are set, e.g. Feynman's vector potential, Born's probability amplitude or Hiley's quantum potential (Davies & Brown, 1999) p138. In physics it is Ψ, the quantum wave amplitude that spreads. In philosophy, it is the virtual reality existence dimension. The challenge of this chapter is to explain the physical behavior of light with only one extra dimension.

---

[9] The imaginary dimension has units *i*, where *i times i* = -1. In normal multiplication, 5 times 4 *repeats* it four times, to give 20. In complex numbers, 5 times *i* *rotates* it by 90° into imaginary space. Multiplying 5 by 4*i* rotates it by 90° four times, to give the original 5 again.

**The Planck program**

Virtual worlds, like Second Life, involve programs, processing, screen nodes and pixels. The processing by a central processing unit (CPU) reads a program to direct screen nodes to set an image's pixel values. This model has no external observer, as the system is observing itself, so grid nodes are both screen and CPU in our terms. Each node transmits instructions like a CPU and also represents a pixel state like a screen. Nodes receive and transmit entity program instructions, which evolve as quantum state pixels by the equations of quantum theory.

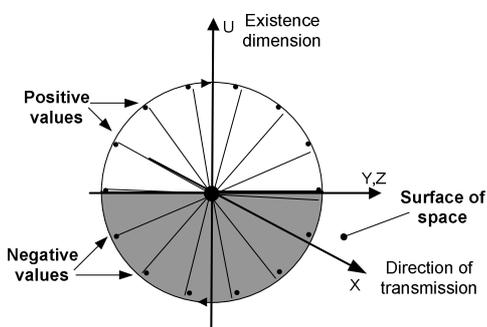

**a.** *Empty space*: A Planck program

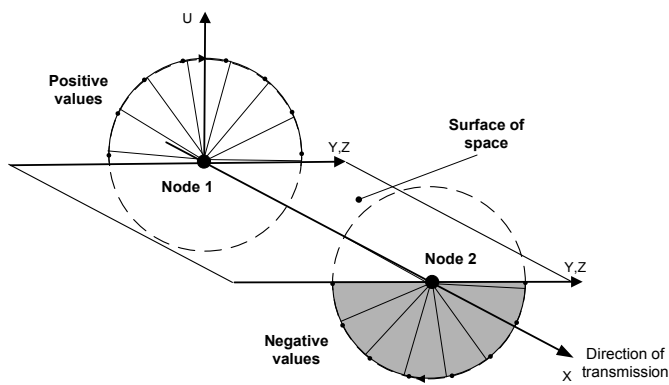

**b.** *First light*: A Planck program splits

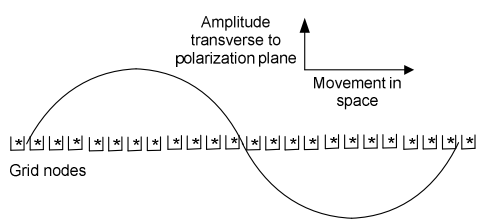

**c.** *Normal Light*: Many nodes setting the basic operation projects a sine wave

*Figure 7. Processing space and light*

In our computing, a CPU *command set* is the basic acts it can apply to current register values, e.g. plus one adds one. The trend to complex instruction set computing (CISC) reversed as reduced instruction set computing (RISC) was found to be more reliable. The command set proposed is the ultimate RISC design, of one command:

> *Set the next value in a circle of values transverse to our space[10].*

This command has the advantage of always applying, as a circle's end is also its beginning. A circle on the surface of space gives absolute positive-negative displacements. Each *click* sets a new value, and a circle of clicks is a *tick*. The set of instructions to add one enough times to complete a full transverse circle is a *Planck program*.

As an analogy, consider a carnival wheel of black-white segments spun by a machine[11] (Figure 7a) where a full wheel turn is one machine cycle. The machine turning the wheel is the grid, the pattern is the program run, a segment is one basic command and the net visual effect "blur" is a quantum state. The machine turns the wheel all or nothing, just as in processing there is no half CPU cycle. If a full pattern is on the wheel, its segments cancel, as equal positive and negative values give a net zero, or as equal up-down displacements produce no effect. So if one machine (grid node) runs a full pattern (Planck program) the result is null processing (empty space).

---

[10] In the next chapter, anti-matter is proposed to arise when the same process runs in reverse.

[11] This mechanical analogy is just to help understanding, and should not be taken literally. There are no actual physical wheels or mechanisms to turn them. In this model everything is information processing.

Now let the same pattern divide over two machines (Figure 7b), where one has the white segments and the other the black ones. If the whole pattern runs, so after white is black and after black white. As each machine now shows something, the effect is no longer null. Something "exists", if only for one cycle. So a Planck program distributed over two nodes can run fully on both over two cycles. Let this represent the highest frequency and shortest wavelength possible for light. In the previous chapter, this *first light* arose when a node of the pristine grid somehow split and "moved" on the inside face of the hole created at the initial event. Also let the patterns be passed along a row of machines each cycle, as the light moves.

The rest of the electro-magnetic spectrum now derives as the expansion of space distributes the same program to more nodes to give longer wavelengths and lower frequencies. As the wavelength increases, more nodes run the same Planck program at a slower rate, giving the sine wave amplitude of light (Figure 7c). The entire electro-magnetic spectrum, from radio-waves to gamma rays, is then a universal Planck program more or less distributed across the grid.

## Energy as processing rate

Energy is a construct useful to physics because it works, but no-one really knows why. It has many forms: kinetic energy, radiant energy, chemical energy, heat energy, nuclear energy, electric energy, magnetic energy, potential energy and by Einstein, mass. In the nineteenth century, higher frequency light was seen to have higher energy, with a ratio as it turned out of the frequency squared. Black body objects absorb and emit light equally at all frequencies, so increasing their temperature should increase the energy of higher frequencies more. This gave what physicists called the ultra-violet catastrophe, e.g. an enclosed furnace is a black body, as radiation bounces around inside it to create every frequency, so a hot furnace should give a fatal dose of x-rays, but in practice it doesn't.

Planck solved this problem by making radiation discontinuous, so atoms can only emit photon energy as a frequency multiple of a basic quantum[12]. As atoms never get enough energy for the highest frequencies, this predicts black body radiation correctly. Einstein deduced from the photo-electric effect that this quantization is a property of light itself, not the atoms, as Planck thought. That light energy is not continuous but comes in fixed packets was unexpected. Why electro-magnetism must be emitted and absorbed in fixed amounts remains a mystery of physics to this day.

In this model, *energy is the node processing rate*. High frequency photons with few nodes in their wavelength must each process at a fast rate. Low frequency photons with long wavelengths have more nodes for the same program, so each processes at a slower rate. The electro-magnetic spectrum is then the same Planck program more or less distributed. Planck's constant is one Planck program per second, a tiny energy, so the energy of a light wave is that times its frequency in cycles per second[13]. Planck's constant. Equally the processing rate per node times the number of wavelength nodes should equal the Planck program rate, as it does if adjusted by the grid refresh rate (the speed of light) to get the right units[14]. A photon's energy then comes in discrete packets because it is processing done by discrete nodes. A higher frequency is one less node to run the same program, giving a discrete energy jump.

If light is the same program that gives empty space distributed on the grid, photons are just space spread out. They have zero rest mass because if a photon rested at a node for its wave train to catch up, it would become empty space. So the frequency of light is always less than space. Equally, higher

---

[12] The word quantum just means "a discrete amount". This amount is now called Planck's constant.

[13] Or $E = h.f$, for energy E, frequency f, and Planck's constant h

[14] For wavelength $\lambda$ and speed of light c: $f.\lambda = c$, so $E = h.c/\lambda$, *giving* $h = E.\lambda/c$

electro-magnetic frequencies are harder to come by, as fewer nodes per wavelength mean bigger energy jumps. The highest possible frequency, with a two Planck length wavelength, must double its energy to reach the next higher frequency, of empty space. Hence it is said that:

"... *vacuum state is actually full of energy...*" (Davies & Brown, 1999) p140.

### The size of space

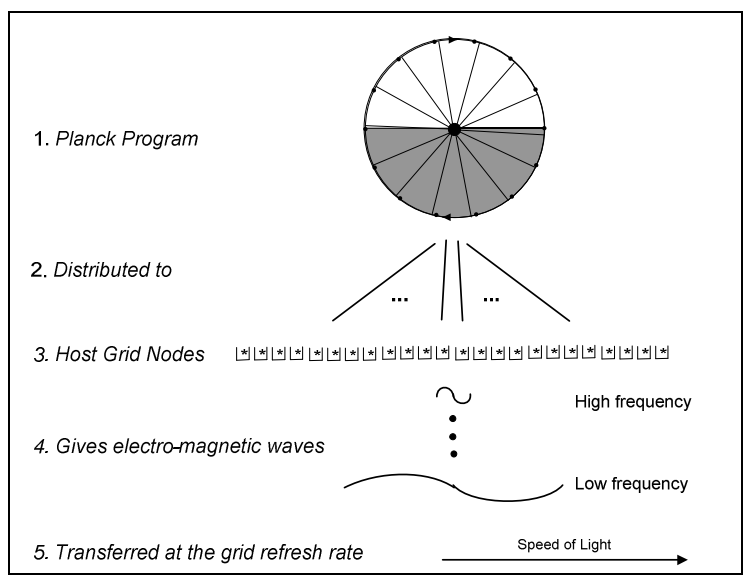

1. *Planck Program*

2. *Distributed to*

3. *Host Grid Nodes*

4. *Gives electro-magnetic waves*
   High frequency
   Low frequency

5. *Transferred at the grid refresh rate*
   Speed of Light

*Figure 8. Deriving the electro-magnetic spectrum*

Plank's constant also defines the granularity of space*:* if it were smaller, atoms would be smaller, and if it were larger, quantum effects would be more evident. Why should the basic unit of photon energy also define the size of space? There seems no reason for the two to connect.

This chapter defines energy as the rate a grid node runs the Planck program, which is a *transverse circle* of values set in sequence. If Planck's constant is this amount of processing per second, it will depend on the number of values in a transverse circle.

The last chapter defined the directions of space at a point by a node *planar circle,* of the neighbors that connect in a planar transfer channel (Whitworth, B., 2010). The number of neighbors in a planar circle defines a circumference, which by Pythagoras's theorem defines a radius, i.e. the "distance" between grid nodes. So the number of grid nodes in a planar circle defines the granularity of space.

If *the grid is symmetric*, transverse and planar circles will have the same number of nodes. So Planck's constant as the size of a transverse circle will also define the size of the planar circle that sets the size of space. Planck's constant links the quanta of energy and the pixels of space because it is the connection density of the grid network that creates both energy and space.

### Delivering a photon's energy

A photon hitting a photographic plate makes one dot but a physical wave's energy should arrive as a smear. It should also take time to arrive as radio wavelengths are many meters long. If a photon as a physical wave has measurable delay from when it first hits to when the rest of the wave arrives, what if it hits something else meantime? The problem is:

"*How can electromagnetic energy spread out like a wave … still be deposited all in one neat package when the light is absorbed?*" (Walker, 2000) p43

In quantum mechanics, a photon delivers *all* its energy *instantly* at a point. This is impossible for a physical wave but not for a processing one. If every node of the photon wave runs the same Planck program, there is nothing to "gather" over its wavelength. Any point can instantly deliver the entire program producing the wave. How this occurs is covered in more detail shortly, but is essentially that a grid node overload causes the photon program to reboot.

## HOW COME THE QUANTUM?

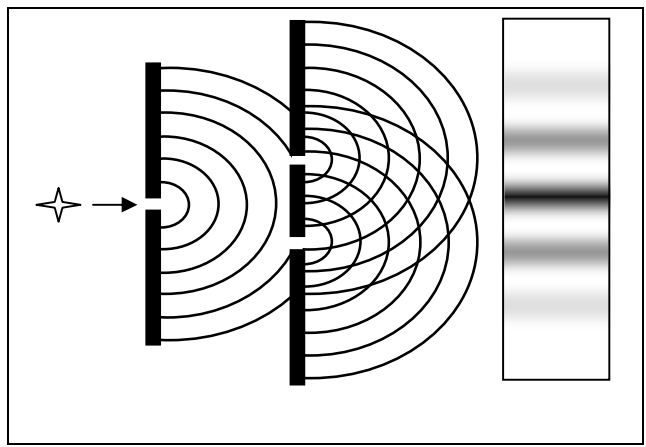

A photon is here a Planck program shared on the grid (Figure 8). The first photon arose as a grid node tore apart in the initial event. For a brief period of inflation others followed suit, causing all the "free" processing that generates this virtual universe. The hole healed but still expanded as space, so light descended into the lower and lower frequencies we call electro-magnetism. It transmits undiminished to this day at the grid cycle rate, which we call the speed of light. The photon Planck program we describe by Planck's constant. So does this rather radical view match how light behaves?

*Figure 9. Young's double slit experiment*

### Young's experiment

Over two hundred years ago Thomas Young carried out an experiment that still baffles physicists today - he shone light through two nearby slits to get an interference pattern on a screen (Figure 9). As only waves diffract like this, a photon must be a wave, but then how does it hit at a point? Or if it is corpuscles, how does it interfere? To find out, physicists sent *one photon at a time* through Young's slits. Each photon gave a dot, as expected of a particle, but then the dots formed into an interference pattern, whose most likely impact point was just behind the barrier between the slits! The effect is independent of time, e.g. shooting one photon through the slits each year will still give a diffraction pattern. As each photon can't know where the previous one hit, how can the pattern emerge? Or if each photon spreads like a wave, how can it hit at a point?

In an objective world one could just check which slit a photon went through, but our world's operating system doesn't allow this. Detectors placed in the slits to see where photons go each just fire half the time. Photons *always* go through one slit or other and *never* through both slits at once. In Nature's conspiracy of silence, a photon is always a particle in one place if we look, but when we don't it acts like a wave in many places. That a photon created and detected at a point can in travel diffract as a wave is like a skier traversing both sides of a tree but crossing the finish line intact (Figure 10).

The problem is simply:

1. *If a photon is a wave*, why doesn't it smear over the detector screen, as a water wave would?

2. *If a photon is a particle,* how can one at a time photons give an interference pattern?

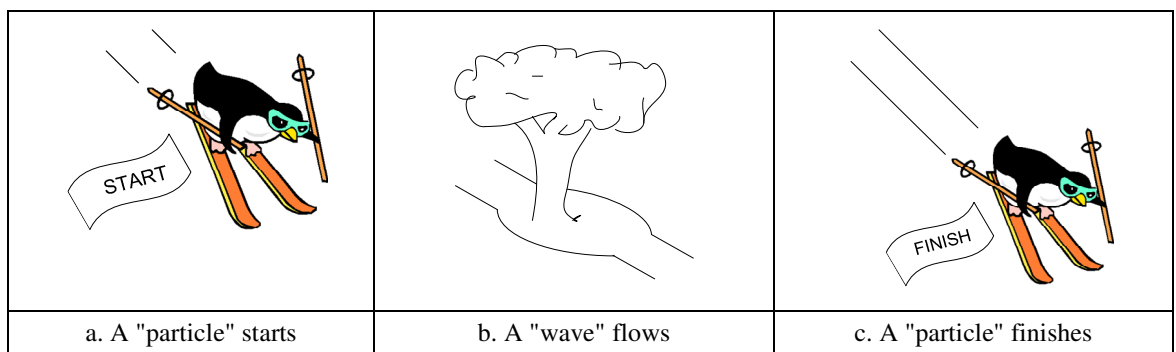

| a. A "particle" starts | b. A "wave" flows | c. A "particle" finishes |

*Figure 10. Wave-particle duality*

Further, this diffraction also occurs for electrons, atoms and even molecules (M. Arndt, O. Nairz, J. Voss-Andreae, C. Keller, & Zeilinger, 1999). Any quantum entity can interfere like a wave in travel, but then appears as a particle when observed. As Feynman says:

"*… all the mystery of quantum mechanics is contained in the double-slit experiment.*"
(Satinover, 2001) p127.

### The Copenhagen compromise

After centuries of arguing whether light is a wave or a particle, Bohr devised the compromise that wave and particle views of light were "complementary", i.e. both true. The truce still holds today:

"*…nobody has found anything else which is consistent yet, so when you refer to the Copenhagen interpretation of the mechanics what you really mean is quantum mechanics.*" (Davies & Brown, 1999) p71.

So if light is both a wave and a particle, physicists can use the appropriate formula as needed. In this *don't ask, don't tell* policy, reality is particles when we look but waves if we don't. Copenhagen enshrined this idea, although everyone knows that a particle isn't a wave and a wave isn't a particle. In no physical pond do rippling waves suddenly become "things" when observed. The "big lie" *that light is a wavicle* has become doctrine, as Gell-Mann noted in his 1976 Noble Prize speech:

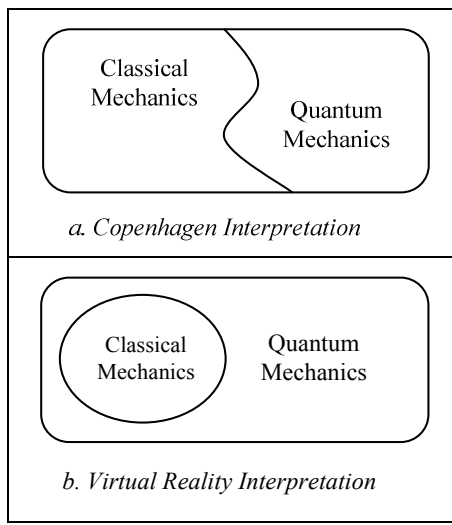

*a. Copenhagen Interpretation*

*b. Virtual Reality Interpretation*

*Figure 11. a. Dual and b. non-dual views*

"*Niels Bohr brainwashed a whole generation of physicists into believing that the problem (of the interpretation of quantum mechanics) had been solved fifty years ago.*"

The mystical wave-particle duality of Copenhagen was a marriage of convenience between irreconcilable wave and particle theories. Like Descartes' mind-body dualism, particle-wave dualism *pretends* that incompatible domains can co-exist separately and equally (Figure 11a). Privately, Bohr denied that the quantum world existed, but held it politic to *suppose* it did to get results. The compromise was necessary, even though quantum theory:

"*… paints a picture of the world that is less objectively real than we usually believe it to be.*" (Walker, 2000) p72.

This model, unlike Bohr, doesn't hedge its bets. In it the quantum world is real and the physical world is derivative, i.e. classical mechanics is a subset of quantum mechanics (Figure 11b) (Audretsch, 2004) p14. If classical physics tells a tale on the stage of physical reality, quantum mechanics is what goes on behind the scenes. A physical photon is here an information transfer, like a screen image when a user clicks. Our observation, or the click, is the long-sought boundary between the classical world we see and the quantum world we don't. Every observation, *by us or anything else*, is a request for information to the quantum database, to generate the virtual world of "things" we see.

### How come the quantum?

Quantum theory explains Young's results as follows: every photon's *wave function* spreads in space, with its power[15] at any point the *probability* it physically exists there. This ghostly wave goes through both slits and interferes as it exits, but if observed suddenly "collapses" to become a thing in one place, as if it had always been so. If we put detectors in the slits, it collapses to one or the other with equal probability. If we put a screen behind the slits, it interferes with itself as it exits the slits then conveniently hits at one screen place, but does so probabilistically according to the interference pattern strength. The mathematics doesn't say what this wave is that goes through both slits, nor why it suddenly shrinks to a point when observed, prompting Wheeler's question: **How come the quantum?**

---

[15] The power of a sine wave is the square of its amplitude.

To see how strange the quantum logic is, suppose the first photon in a two slit experiment hits a screen at a certain point, becoming the first dot of what will *always* be an interference pattern. Now suppose the first photon of another experiment, with a detector blocking the other slit, goes through the same slit to hit the screen at the same point. Just suppose. This is now the first dot of what will *never* be an interference pattern. The difference between these outcomes *must* exist in their first events, yet they are the same – the same photon goes through the same slit to hit the same screen point. For each, whether the slit it *didn't go through* was blocked decides the physical pattern it is part of. If the photon could have gone through the other slit, there is interference, but if it couldn't there isn't. How can a *counterfactual event,* which could have happened but physically didn't, change a physical outcome?

Yet this theory, of imaginary waves that conveniently collapse when viewed works brilliantly. It is the most successful theory in the history of science. Yet it leaves two key issues unresolved:

1. *What are quantum waves?* What exactly is it that spreads through space as a wave?

2. *What is quantum collapse?* Why must the wave collapse if viewed?

Until it answers these questions, quantum theory is a recipe without a rationale and physicists are mathematical witch doctors using a herbal remedy they can't explain.

## What are quantum waves?

As copying information takes nothing from the original, one expects a virtual world to use this feature to advantage. The quantum no-cloning theorem says that we cannot copy quantum states, as to read quantum data is to alter it irrevocably (Wootters & Zurek, 1982), but *the system* that creates those states can by definition easily copy them. The grid, it is now proposed, is the ultimate copying system.

### Distributing quantum processing

Let a Planck program distribute to its wavelength grid nodes as parallel supercomputing distributes a program to many processors. Also, add a *conservation of processing* principle that each instruction can only be allocated once to the grid at a time. Many nodes can then share the same code, e.g. if one man with a shovel digs a hole in one minute, two men sharing the shovel will each dig half a hole in that time. So a Planck program distributed between two nodes will run half as fast on each. In general, *a program distributed runs slower not less at each point.*

In computing, programs are shared by instantiation, an object orientated system (OOS) method that lets information objects dynamically inherit code from a source class. Screen buttons instantiating the same class blueprint look the same because they run the same code. In addition, any code changes immediately reflect to all instances. Let a photon program run instances on many nodes, where each runs the entire Planck program in sequence, but shares the same code with others. If a node can't finish the program in one of its cycles, it carries it on to the next. The photon sine wave now arises as each wavelength node slowly runs the same program at a different phase. As a new node starts at the wave head, another finishes at its tail, so the total processing involved is constant. A photon program divides its existence across space by distributing its processing over grid nodes. If the entity program always allocates all its instructions to the grid, *it always fully exists.*

### Sharing processing

In addition, each node shares its processing with its neighbors each cycle. So the wave not only moves forward on its transmission axis, but also spreads outwards in all directions. This also occurs by instantiation, as above. The grid spreads the processing given it like a pool surface gives ripples when a pebble is dropped on it (Figure 12), except the grid ripples are 3D.

This is the proposed origin of Huygen's principle, that every photon point is also a wave source. The photon wave front then ensues as Huygens proposed, by reinforcement and interference, but of processing not matter. The photon is a processing event reverberating on a grid network. For a pebble drop, the initial energy spreads to larger circles. The *total energy flux* per ripple is the same, except for friction, as Gauss noted. Grid ripples have no friction so the *total processing flux* per sphere surface is constant, i.e. decreases as an inverse square. Physical waves reduce amplitude as they spread but *processing waves just run slower*. If quantum processing spreads on the grid like a 3D spherical ripple, it will:

1. Add at any node point,

2. Decrease as an inverse square with distance, and,

3. Propagate at one grid node per local node cycle (light speed).

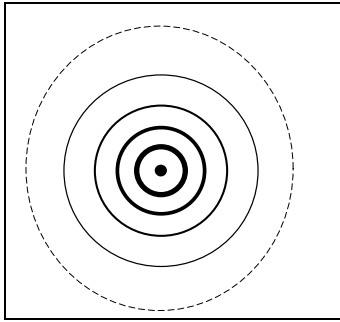

*Figure 12. Pond ripples*

This accounts for electrical, magnetic and gravitational fields that add by field combination[16], decrease in strength as an inverse square of distance, and propagate at the speed of light, e.g. if the sun suddenly disappeared, it would be eight minutes not just before the earth stopped receiving its light, but also before it ceased to feel the sun's gravity.

### *What is existence?*

To Einstein a photon was a physical thing located in space, with physical attributes that defined its motion or rest. A photon that hit a screen at a point had to take a physical path through one of the slits. Its initial state had to define its end state, i.e. where it hit the screen. If its trajectory was unknown, as quantum theory holds, then photons had to have "hidden variables":

*"This is the fundamental problem: either quantum mechanics is incomplete and needs to be completed by a theory of hidden quantities, or it is complete and then the collapse of the wave function must be made physically plausible. This dilemma has not been solved until today, but on the contrary has become more and more critical."*(Audretsch, 2004) p73

Here, quantum mechanics is not incomplete or physically plausible but complete and non-physical. It is not quantum theory that must be made physically plausible, but physicality that must be made implausible. Quantum theory challenges the naïve realism of objective reality, because physical objects can't divide their existence as quantum waves do.

Yet if a photon moves as distributed program instances, which one is *the* photon? The question assumes a photon is one thing, but as will be seen, the answer is any. We only really know that photons interact in one place - that they previously existed so is a conclusion tacked on to the facts. Quantum theory's statement that photons *travel* as probability waves but collapse to *interact* in one place doesn't contradict observation. The critics of quantum mechanics could not fault this logic because there is no fault. In general, to say that an electron *has* a quantum wave function is just the stubborn illusion of inherent physical things. In this model, the electron *is* the quantum wave function.

### Quantum collapse

Quantum mechanics gives no reason for quantum waves to suddenly "collapse" when observed:

*"After more than seven decades, no one understands how or even whether the collapse of a probability wave really happens."* (Greene, 2004), p119

---

[16] If charge 1 has electric field $E_1$ and charge 2 has electric field $E_2$, the electric field at any point $E = E_1 + E_2$

Quantum mechanics formalizes how photon programs spread on the grid as three-dimensional waves. A Planck program photon distributed can't overload a grid whose channel bandwidth is a Planck program[17], but meeting nodes processing the matter of a photographic plate can. A photon arrives at a detector screen as a cloud of packet instances, seeking processing from nodes already busy with the screen's matter. The grid overloads, and if a processor overloads it reboots, i.e. reloads its processing from scratch. So a grid node overload will try to re-read *all* the processing involved in the overload. The *one* grid node that succeeds, and reboots the *entire* photon program, is where it "hits" the screen.

In general, programs on a network "collide" if they overload a node. The reboot re-allocates the processing of both programs in potentially new ways (to us interaction outputs). The only condition is that the total processing before and after is the same. In computing terms, both programs stop and new ones are formed. Now while stopping an instance doesn't affect its parent program, stopping a program template must stop its child instances, as they now have no program source. So if a photon program can spreads instances across many nodes, a reboot in any one restarts the photon program in that node. If one node acquires <u>all</u> a photon program's instructions for one cycle, all the other instances must "disappear". In this case, a reboot causes a spreading program to restart at a node point.

The collapse of the wave function is then the inevitable disbanding of instances when a program reboots. Although the wave seems to disappear, no processing is lost when the whole program resets. Quantum collapse is irreversible because a reboot loses all previous information. The overload won't repeat if nodes *first* share processing with their neighbors, *then* do any instructions received. The reboot then spreads the overload back out to the grid.

To recap, entity programs collide if their spreading instance packets overload the grid, causing nodes to reboot and re-read all processing. The first to succeed is where the entity programs restart from. The grid continuously annihilates and creates quantum entities, but the total processing remains constant, e.g. if two electrons collide we see those leaving as continuing those that entered, but we don't know that, as one cannot mark one electron from another. Here, the electrons leaving are brand new ones, just off the quantum press. In this model, every quantum collapse destroys and creates quantum entities. That particles continuously exist is an impression created by the conservation of processing.

### *Non-locality on the universal screen*

To Einstein, quantum collapse implied faster than light travel so was absurd. In his thought experiment, a photon travels through a slit to hit a screen, so before it hits, by the wave function it could exist at points A or B on the screen with some probability. After it hits, it is suddenly entirely at point A say and not at point B at all. As the screen moves further away, the wave projection increases until eventually the A to B distance is light years, but the quantum collapse is still immediate. The moment point A "knows" it is the photon then B "knows" it is not, even if they are in different galaxies. The collapse decision is applied faster than the speed of light, which is impossible for a physical message. Quantum collapse, like a feather in New Zealand tickling a physicist in New York, contradicts physical reality. How can a decision in one place *instantly* affect another *anywhere in the universe*?

Now suppose a photon is an entity program. A program can instantly change any pixels on a screen wherever they are. The program-pixel connection ignores screen location, i.e. is "equidistant" to every screen point. It doesn't "go to" a screen node to change it, but acts immediately on it. Similarly, a photon program connects directly to instances anywhere in the universe. If in quantum collapse it stops, so must every instance in the universe. As programs ignore screen limits, so quantum entities ignore node-to-node relativity restrictions.

---

[17] By the last chapter, every node-to-node transfer involves a planar channel whose bandwidth is one transverse circle of values, i.e. one full Planck program per cycle.

*The quantum existence lottery*

If a photon instantiation cloud overloads many screen nodes at once, which one reboots? The photon is envisaged as a set of instructions distributed among nodes, which actively share with their neighbors each cycle. The entity program runs its code all the time, to exist, but can't issue the same instruction twice. Equally, each node channel runs the instructions it receives up to its finite bandwidth.

A network server must run faster than its client workstations. To a user typing, a client computer is fast, but in-between user key presses a server can handle hundreds of other clients. To the server, the client is slow, so a node cycle fast to us must be slow to an entity server[18]. If an entity program serves many nodes that reboot, each one could:

1. *Have program access.* In this case, it reads all the program instructions that cycle, denying all other instances access. This is then to us a physical event.

2. *Have no program access.* In this case, it tries to read the program but gets a "busy" response, so drops the instance. This is then a potential event that didn't happen.

A photon processing cloud overloading the grid is a winner takes all lottery. The first node to successfully lock program access wins the prize of *being* the photon, by reading all its code that cycle. As other instances can't get access that cycle, they are dropped. The photon program generate a legion of instances, and can be "born again" from any one at any time.

## The quantum probability of existence function

When a photon quantum wave meets a detector screen, quantum theory specifies the probability a quantum wave will collapse to a point as follows: First, Schrödinger's quantum wave equation gives all the values at a point (a spreading wave can reach the same point by more than one path). Then cancel positive and negative values to give a total amplitude. Finally, square that for the probability it is there.

In this model, if many nodes overload, which will reboot successfully? If a program services many client nodes, the first reboot request received will lock out any others. If a photon wave overloads a screen at many points, the restart node will be the one with access to the program at the time. Nodes running more instructions from that program will get more access to it, so are more likely to reboot successfully. So the reboot probability should vary linearly with the number of photon instructions read.

To estimate this, first define the photon instances a node gets, then cancel positive and negative calls before requesting processing, as a processing efficiency. The resulting total processing requested is the probability the node will reboot successfully. The amplitude quantum mechanics squares is a one dimensional projection of a two dimensional processing wave. A projection is a dimension reduction, as three dimensions project a shadow on a flat surface. If the quantum amplitude we measure projects a 2D process, the processing done will be its square, as a sine wave's power is its amplitude squared.

Quantum theory's probability of existence is the amount of entity processing done at a node. It squares the quantum amplitude because processing has a complex dimension. Grid nodes that process more have more program access, so reboot successfully more often. The outcome is random *to us* as it involves program-node services we have no access to. That processing waves spread, add and collapse as quantum waves do suggests that the physical world is a quantum processing output.

*Summary*

Table 1 interprets Feynman's principles of quantum mechanics as a network protocol to resolve packet collisions (Feynman et al., 1977) p37-10. In Young's experiment, a photon program divides its processing into instance packets that spread, as per Huygens, to go through both slits and interfere as they exit. At the screen, many nodes overload and request a program reboot. The first one to succeed is

---

[18] Entity program access time must be less than Planck time, estimated at about $10^{-43}$ seconds.

where we see the photon "hit" the screen. This follows the interference pattern observed because it varies with the net processing done, even for one photon at a time. If detectors are placed in both slits, a reboot in either will occur. If a detector is placed in one slit, it will only fire half the time, as half the time the photon will "exist" in the open slit, and there will now be no diffraction effect.

*Table 1. Quantum mechanics as a network protocol*

| Quantum theory | Network protocol |
|---|---|
| *1. Existence.* The probability a photon *exists* is the absolute square of its complex probability amplitude value at any point in space[19] | *1. Reboot.* The probability a node overload *reboots* a photon program successfully is the processing done, which is the absolute square of its amplitude |
| *2. Interference.* If a quantum event can occur in two alternate ways, the positive and negative probability amplitudes separately combine at every point, i.e. they interfere[20] | *2. Combination.* If program entity instances arrive at a node by alternate grid paths, positive and negative values cancel in every grid node, i.e. they interfere |
| *3. Observation.* Observing one path lets the other occur without interference, so the outcome probability is the simple sum of the alternatives, i.e. the interference is lost[21] | *3. Obstruction.* An obstacle on any path obstructs instances traveling that path, letting the alternate path deliver its processing unchanged, i.e. the interference is lost |

Quantum theory is to us strange because it describes how processing creates the world we see as an objective reality. We see photon programs *interacting* in one processing place, so how can they *transmit* as distributed instance packets? The processing cloud that is the quantum wave function is only physically "a photon" when it restarts at a node reboot. It moves as distributed processing but arrives as a program restart. So if one asks if it goes through both slits at once, the answer is yes. If one asks if it arrives at one point on the screen, the answer is also yes. What travels as a spreading wave arrives as a local "thing", and we see only the latter.

## TAKING EVERY PATH

Newton rejected Huygens's wave theory of light because:

"*For it seems impossible that any of those motions … can be propagated in straight lines without the like spreading every way into the shadowed medium on which they border.*" (Bolles, 1999) p192

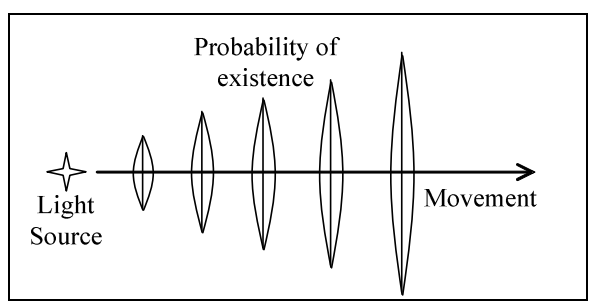

*Figure 13. The probability of existence of light*

If light is a wave spread out, he argued, how can it travel in straight lines through optical devices? Sound is a wave that bends around corners, allowing us to hear people talking in the next room. In 1660 Grimaldi showed that light also spreads out, but less so as it has a shorter wavelength. Figure 13 shows how a photon wave could vary in power along its line of travel, where the photon is more likely to exist at the thicker sections. The probability maxima are a straight line, but at each stage it is spread out. This predicts what is found, as if a single photon is detected by a series of screens at different distances, the hits are not in a perfect straight line, but randomly distributed about it (Figure 14). Particle theory would need photons to travel in a zigzag path to explain this.

---

[19] If U is the quantum wave amplitude, and P its probability, then $P = |U|^2$ for one channel.

[20] If $U_1$ and $U_2$ are the probability amplitudes of the two ways then the total amplitude $U = U_1 + U_2$. If $P = |U_1 + U_2|^2$, then $P = P_1 + P_2 + 2\sqrt{P_1 P_2} \cos(\theta)$ where the latter is the interference for phase difference $\theta$.

[21] Now $P = P_1 + P_2$ with no interference term.

### The law of least action

If a photon wave travels in a straight line *on average,* why can't it sometimes "bend into the shadows", so we see a torch beam from the side? Why doesn't it have a sideways wake behind it, like the turbulence of a high speed bullet? Behind the problem of how a spreading wave becomes a ray of light lies a deeper one that has puzzled thinkers for centuries. Hero of Alexandria first noted that light always takes the shortest path between points, raising the question of how it knows that path? It might seem obvious it is a straight line, but how at each point does a photon know what "straight" is?[22]

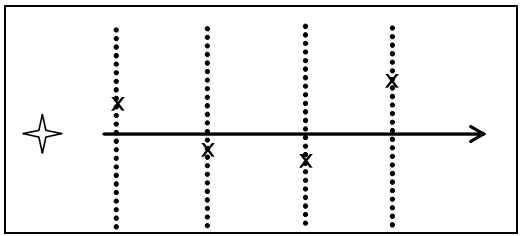

*Figure 14. Detection of a "ray" of light*

In 1662 Fermat amended the law to be the path of least time, as light refracts to take the fastest path not the shortest one. Refracted light changes direction as it enters a transparent medium like water, where it travels slower (Figure 15). Imagine the photon as a life guard trying to save a drowning swimmer as quickly as possible. Is the dotted straight line shown the quickest path to the swimmer? If a lifeguard runs faster than he or she swims, it is quicker to run further down the beach then swim a shorter distance, as shown by the solid line in Figure 15. The dotted line is the shortest path but the solid line is the fastest, and that is the path light takes. Again, how does it know *in advance?*

In 1752 Maupertuis developed the general principle that:

"*The quantity of action necessary to cause any change in Nature always is the smallest possible*".

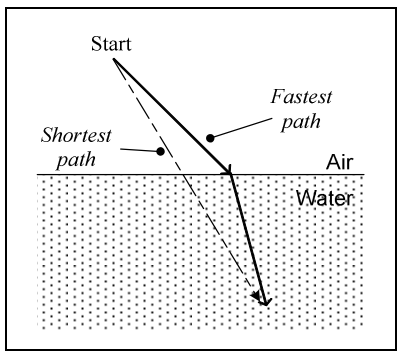

*Figure 15. Wave refraction*

This *law of least action,* that light always takes the most efficient path, was developed mathematically by Euler, Leibnitz, Lagrange, Hamilton and others, sparking a furious philosophical debate on whether we live in "the best of all possible worlds". Despite Voltaire's ridicule, how a photon finds the best path remains a mystery today, e.g. light bouncing off the mirror in Figure 16 could take any of the dotted paths shown, but by the principles of optics always takes the solid line fastest path. How does it actually do this? As the photon moves forward in time to trace out a complex path, how does it at each stage pick out *in advance* the shortest route? Few can see the problem, but as Feynman says:

"*Does it 'smell' the neighboring paths to find out if they have more action*?" (Feynman et al., 1977) p19-9

To say that a photon chooses a path *so that* the final action is the least gets causality backwards. That a photon, the simplest of quantum entities, with no known internal mechanisms, always takes the fastest route to any destination, for any combination of media, for any path complexity, for any number of alternate paths and inclusive of relativistic effects, is nothing short of miraculous.

### The law of all action

Feynman proposed that photons actually take all possible paths, as if they can't, all paths become equally likely (Feynman et al., 1977) p26-7. His *sum over histories* method supposes that light goes from A to B by all possible paths then chooses the one with the least action integral. It exactly predicts light travel but is physically impossible, so by the Copenhagen view is fictional. Yet in this model, the photon does exactly this: *photon instances take all available paths of the underlying grid architecture.* Physical reality is then decided by those that find the fastest route to trigger a detector overload. That reboot then destroys all other instances, like a clever magician removing the evidence of how a trick is done. The photon wave takes all paths and its first restart *becomes physical reality*, complete with path taken. In computing, leaving decisions to the last moment is called just in time (JIT) processing. So Feynman's theory isn't fiction: the photon really does take all paths. Physical reality selects the fastest path to a detector on a first come first observed principle.

---

[22] In relativity light doesn't always travel in a straight line, so "straightness" is not self-evident.

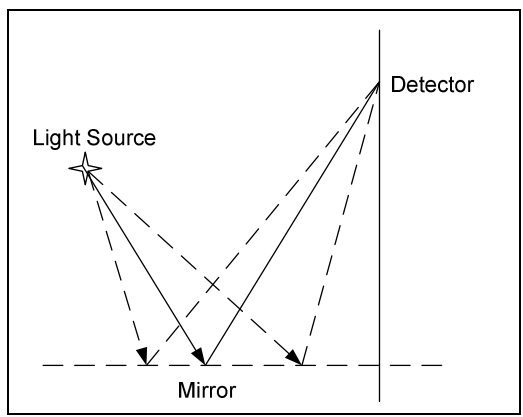

*Figure 16. Principle of least action*

Indeed, how else could the law of least action arise? A photon *cannot* know in advance the fastest route to an unspecified destination *before* it leaves. It must take all possible routes and let the system holistically choose physical reality *later*. Taking all routes might seem an inefficient way to travel, but in a virtual world calculating a path and taking it is the same thing. The system must calculate all possible paths anyway, so taking them all then picking the best is the obvious option.

So how does a light photon, the simplest of all things, always know in advance the best way to any destination? It doesn't! A photon is here a spreading wave taking all possible paths. If it "hits" a detector point it re-spawns there, with quantum collapse the necessary garbage collection of other instances. We call that physical reality. What chooses it isn't the restart instance but the system exploring every path. What appears as an *individual event* done by one photon is really a *holistic event* achieved by a throng of instantiations.

We assume that the physical reality we see is all there is, but here it is the end product of a great deal of unseen quantum processing. If entity instances spread by all possible grid channels, then all that can happen does happen, but just not physically. In this "evolutionary physics", quantum processing calculates all the options and physical reality takes the best and drops the rest. The physical principle of least action implies a *quantum principle of all action*, that:

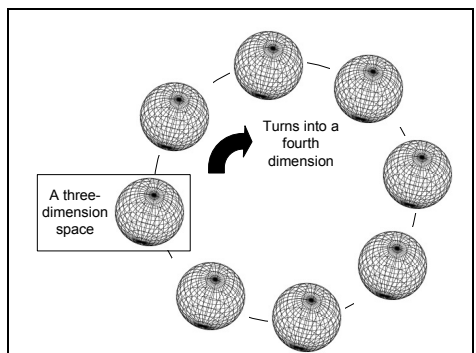

*Figure 17. Adding a fourth dimension*

> *Everything that can occur in physical reality does occur in quantum reality*[23].

If physical reality happens after quantum reality tries every option, if this isn't the best of all possible worlds, it isn't for lack of trying.

### QUANTUM SPIN

Light as an oscillation outside space explains its spin.

### Vibrating the fourth dimension

Adding a dimension to 3D space doesn't just add another direction, it turns *all of space* into that dimension. So a sphere gives a hyper-sphere, or sphere of spheres (Figure 17)[24]. While 3D space has three perpendicular planes, 4D space has six planes perpendicular to each other[25]. Adding a dimension doubles the number of perpendicular planes. The three extra planes are not just perpendicular to our space, but also to each other, i.e. *quantum amplitudes outside our space can be orthogonal to each other.* It is hard to grasp but mathematically defined that a quantum direction is perpendicular to a plane in our space, not a point. Each point in our space has three orthogonal quantum amplitudes, at right angles to the three orthogonal planes through it. So light can vibrate in three independent amplitude directions at every point on the hyper-surface we call space.

---

[23] Or as Feynman states, "Whatever is not explicitly forbidden must happen". Gellman called it the quantum totalitarian principle.

In physics, a photon's electric field is a complex value rotating in an imaginary dimension, which here actually exists. The sine wave a photon presents to us is a rotation in quantum space moving in our space, whose amplitude is transverse to the photon polarization plane (Figure 18)[26]. A photon moves in space but rotates outside it at right angles to its polarization plane. *The polarization of light is the plane in our space perpendicular to its quantum amplitude.*

So a polarized filter blocks polarized light in quantum space, not physical space. One filter can block vertically polarized light, which vibrates outside space at right angles to the vertical plane, while another is needed to block a horizontally polarized light, which vibrates outside space at right angles to the horizontal plane. These vibrations are at right angles not only to our space *but also to each other*, i.e. vertically polarized light can pass through a filter that blocks horizontally polarized light.

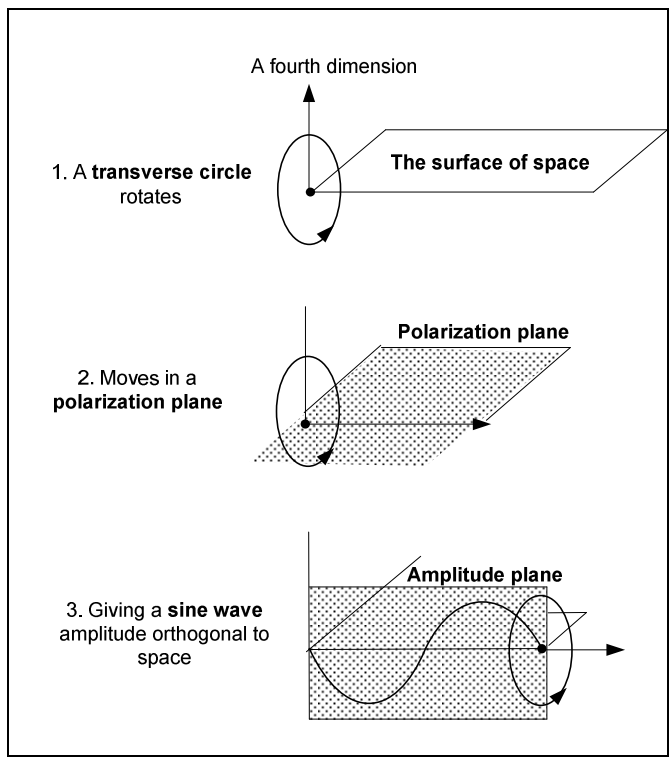

*Figure 18 A photon gives a sine wave*

### Quantum structures spin in space

A light filter set at an angle to light's polarization plane still lets some photons *entirely* through. A greater angle lets fewer photons through, but it is still all or nothing, e.g. an 81° filter lets 10% of photons through, which exit polarized at the filter angle. How can a photon pass entirely through a filter that nearly blocks it entirely? Let the photon *spin around its axis of movement,* as a bullet spins as it flies through the air. For a structure to spin it needs a:

a. *Rotation axis.* Around which the spin occurs. This dimension doesn't change with the spin.

b. *Rotation plane.* In which the spin occurs. These dimensions swap values as structure spins.

For example, a spinning propeller from the front displays its rotation plane. As each blade turns, its vertical and horizontal extents swap. Viewing the propeller from the side shows only one rotation plane dimension, the vertical. Now the propeller blade seems to appear and disappear, as the spin swaps its vertical and horizontal extents, but really it is turning into an unseen horizontal dimension.

---

[24] If physical space has dimensions (X,Y,Z), quantum space has dimensions (X,Y,Z,U), where U is a fourth, unseen dimension. Our space is just a surface in quantum space.

[25] Physical space with three dimensions X, Y and Z has three planes XY, XZ and YZ. Quantum space with an additional U dimension has three additional planes XU, YU and ZU.

[26] By "moving" is really meant "transmitted".

Spin in four dimensions works like spin in three, but with more options. So a 3D structure in a 4D space can turn on an axis outside itself, as can 2D structure in a 3D space. A photon is here a three-dimensional structure with a movement dimension (X), a polarization plane (XY) and an unseen dimension (U) into which it rotates (Figure 18). This leaves a free space dimension (Z) for it to *spin around its axis of movement*. It then spins into all other possible polarization planes (Figure 19).

Its quantum amplitude direction doesn't change as it spins, because it isn't on the rotation plane[27]. So as a vertical polarized photon spins into a horizontal plane, its quantum existence "disappears", as the propeller blade did when seen from the side. The effect is continuous as the polarization plane angle increases[28]. It is like turning a thin piece of paper held up: first it is visible, then edge on it can't be seen, then it re-appears again.

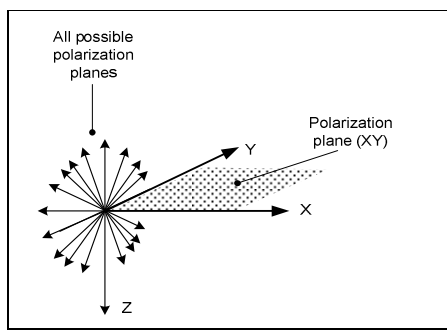

A photon *spins once per grid node cycle* to simultaneously exist in many planes. It's processing existence projects into the planes cutting its movement axis according to angle. Processing collisions are all or nothing, so observing it delivers the entire photon program to a reboot, restarting it polarized in that plane. A photon randomly goes entirely through a filter for the same reason it randomly hits a screen point – any reboot of its distributed existence delivers the entire entity program.

*Figure 19. Polarization planes*

### What is quantum spin?

In classical spin, an object extended in space, like the earth, spins on an axis into a rotation plane (Figure 20). Measuring spin on any axis gives some fraction of its total spin, and measuring spin on three orthogonal axes gives its total spin. In contrast, quantum spin on *any axis* we care to measure is always a multiple of Planck's constant. It is also in either direction randomly and if measured on one axis can't be re-measured on another. Indeed, quantum spin is so strange that when Pauli first proposed it, he was not believed:

> "… the spin of a fundamental particle has the curious feature that its **magnitude** always has the **same** value, although the direction of its spin axis can vary…" (Penrose, 1994) p270

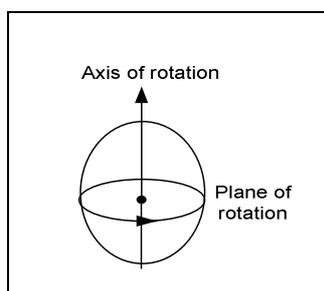

A photon exists outside space but spins in it, so its quantum amplitude varies as it spins, by the nature of four dimensional space. This alters the probability of successfully observing a photon in a plane, but not its effect, which is always the full program. Just as a photon can entirely pass a filter on an angle, so successfully measuring its spin on any axis always gives all its spin. The spin delivered is that of a transverse circle, or Planck's constant expressed in radians[29].

*Figure 20. Classical spin*

A photon's spin direction is like its movement direction, as it is just transmits down all possible grid architecture channels. It can divide its processing in clockwise and anti-clockwise directions, then if a particular instance restarts take that spin only. The event can't be redone because all previous information is gone. Imagine asking which way a coin is spinning on a table if it is too fast to see. One can only know by stopping it, which then can't be repeated, unless you re-spin the coin, which is a new case. The quantum coin spins both ways at once and at every point on the table! In the next chapter, complex electron structures "half-spin" in the physical world.

---

[27] The Planck transverse circle already turns around the X axis into the YU plane, but the photon can still spin in the YZ plane. This swaps its Y and Z values while leaving U and X unchanged. U remains perpendicular to XY, so as Y and Z swap it becomes invisible, as it has no extension orthogonal to the XZ plane.

[28] If U is the original existence it reduces as U.Cos($\theta°$) where $\theta°$ is the angle from in the original plane. So at a 90° angle it has no value as Cos(90°) = 0.

[29] Spin is actually expressed in Plank's reduced constant of $\hbar$ (h-bar) = h/2$\pi$ (in angular radians).

## PHYSICS REVISITED

This model suggests answers to some of the perennial problems of physics.

### Superposition

In mathematics, solving an equation gives solutions that satisfy its conditions. Solving the quantum wave equation also gives solution "snapshots" of possible results, each with an associated probability. These evolve dynamically over time, forming at each moment an orthogonal ensemble, only one of which can physically occur. This mathematics has an unusual feature: if any two states are solutions, so is their linear combination[30]. While single state solutions match familiar physical events, combination states never physically occur, yet they underlie the mysterious effectiveness of quantum mechanics. Quantum combination states behave quite differently from physical states, e.g. it is in such a combination or *superposition* state that one photon goes through both Young's slits at once.

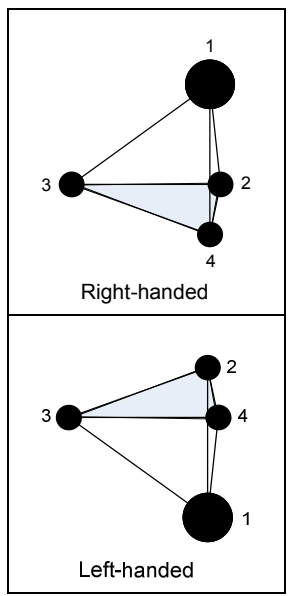

*Figure 21. Ammonia molecule states*

For example, ammonia molecules have a pyramid shape, with a nitrogen atom apex (1) and a base of hydrogen atoms (2,3,4), as in Figure 21, and manifest in either right or left handed forms (Feynman et al., 1977) III, p9-1. To turn a right-handed molecule into a left-handed one, the nitrogen atom must pass through the pyramid base, which is physically impossible. Yet in quantum mechanics, if two states are valid then so are both at once. So observing an ammonia molecule finds it left-handed one moment and right handed the next, yet it can't physically oscillate between these states.

Superposition isn't just ignorance of a hidden physical state, as the molecule can exist randomly in physically incompatible ways at once. To see quantum superposition as adding classical states is to misunderstand it, e.g. superposed electric currents can flow both ways round a superconducting ring at the same time, but physically such currents would cancel (Cho, 2000). Only if measured does one or the other physical state manifest. Here, superposition is a quantum entity program distributing its processing existence down all possible grid channels. If photon instances spin clockwise and anti-clockwise at once, in superposition, it "half-exists" in both spins. This contradicts our idea of objective reality but is just business as usual in the quantum world.

### Schrödinger's cat

Schrödinger found quantum superposition so strange he tried to illustrate its absurdity by a thought experiment. He imagined his cat in a box where photons randomly radiated might hit a detector, to release a deadly poison gas. In quantum theory, photons both exist and don't exist until observed, so as photon plus detector is also a quantum system, it also superposes in a detected and undetected state. By the same logic, the poison release system also both fires and doesn't fire, so the cat is in an alive-dead superposition, until Schrödinger opens the box.

If a photon can both exist and not exist, can Schrödinger's cat be both alive and dead? Or if cats can't be alive and dead, how can photons both exist and not exist? Or if photons can superpose but cats can't, as quantum entities form classical entities, when does the superposition stop? In this model, there is no infinite observer regress. Quantum collapse follows *any* grid node reboot, so the superposition

---

[30] If $\Psi_1$ and $\Psi_2$ are state solutions of Schrödinger's equation then ($\Psi_1 + \Psi_2$) is also a valid solution

uncertainty doesn't cumulate beyond the first node reboot. Schrödinger may not know if his cat is alive or dead, but *the cat does*.

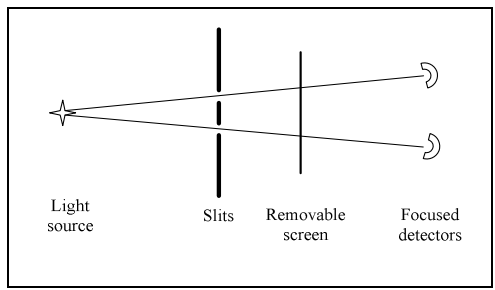

*Figure 22. Delayed choice experiment*

### Retrospective action

That photons travel at about a foot per nanosecond allows a *delayed choice* two slit experiment, where detectors are turned on *after* the photon passes the slits. Two detection options are set up. The first is the usual screen but in the second this screen is quickly removed to reveal two telescopes that focus on only one slit or the other (Figure 22). The experimenter chooses the screen or telescope set-up *after* the photon passes the slits. The screen then gives interference, so the photon took both paths, but if the screen is removed to reveal the telescopes, only one fires, so the photon took one slit or another. The inescapable conclusion is that detectors turned on *after* the photon passed the slits decide its path:

"*Its as if a consistent and definite history becomes manifest only after the future to which it leads has been settled.*" (Greene, 2004) p189

That an observation made *after* a photon travels decides the path it took *before* it was observed is backward causality – the future affecting the past. And the distances involved are irrelevant. A photon could travel from a distant star for a billion years and decide when it arrives at earth if it "actually" came via galaxy A or B. As Wheeler observes:

"*To the extent that it {a photon} forms part of what we call reality... we have to say that we ourselves have an undeniable part in shaping what we have always called the past.*" (Davies & Brown, 1999) p67

For an objectively real physical world this experiment shows that time can flow backwards, which denies causality and puts into doubt much of physics.

In contrast, let photon instances travel all grid paths and physical reality be generated at the detector on demand. The photon program only becomes *a physical photon* and its path *the path the photon took* when it restarts. Adding or removing detectors en-route makes no difference if the wave takes every path anyway. It isn't retrospective action if physical reality is created when the quantum wave arrives at a detector. A photon can take all paths until disturbed then appear at that point complete with a physical path history. This model saves physics from reverse causality.

### Non-physical detection

A Mach-Zehnder interferometer, a device originally designed by John Wheeler, can detect an object on a path not traveled. In Figure 23, a light source shines on a beam splitter which sends half its light down *path 1* and half down *path 2*. Path 1 has a mirror pointing to detector 1, and path 2 has a mirror pointing to detector 2. At this point, the light travels both paths equally and each detector fires half the time. Now add a second beam splitter where the two paths cross, to again send half its light to each detector. With this splitter in place, light is only detected at detector 1, never at detector 2. Sending one photon at a time through the system has the same effect - detector 1 records it but detector 2 doesn't respond at all.

Quantum mechanics explains this by fictional quantum states that evolve down the two paths, with each mirror or splitter turn of direction delaying their phase. The paths to detector 1 have two turns, so states traveling them phase shift by the same amount, but path 1 to detector 2 has three turns while path 2 has only one. Paths phase shifted this way cancel out, so nothing is seen at detector 2.

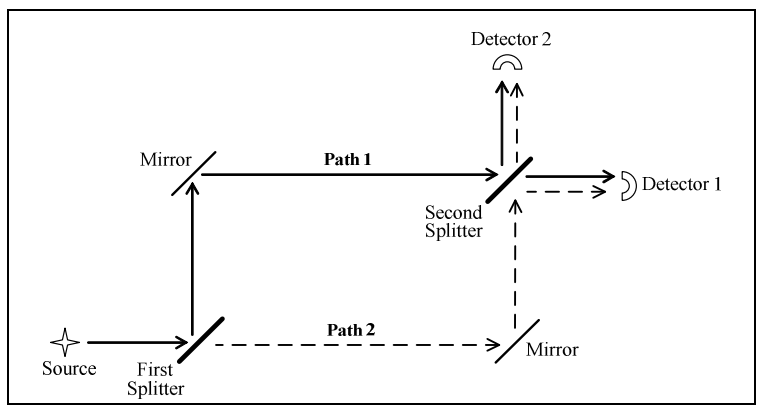

*Figure 23. The Mach-Zehnder interferometer*

Remarkably, this setup can register an object *without physically touching it* (Audretsch, 2004) p29. Suppose that on path 2 is a bomb so sensitive that even a single photon will set it off. Shining even a photon of light on it, to see if it is there, explodes the bomb. However if the bomb is placed on path 2 of a Mach-Zehnder interferometer, the usually quiet detector 2 will sometimes respond, *without exploding the bomb*. As this never happens if path 2 is clear, so *proves* a bomb is on the path, though no light has touched it physically. This bomb detection technique however sets the bomb off half the time. This non-physical detection has been verified experimentally, though not of course with bombs (Kwiat, Weinfurter, Herzog, Zeilinger, & Kasevich, 1995). For light shone through a Mach-Zehnder interferometer:

1. With two clear paths, only detector 1 ever fires.

2. If a receptor sensitive to *any* light is put on a path, the silent detector now sometimes fires.

3. This occurs *only* with a receptor on that path, which physically registers nothing.

In this model, the states of quantum mechanics actually occur. Photon program instances travel all four paths to both detectors with equal probability (Table 2). If both paths are clear, the instances reaching detector 2 by both paths interfere, so it never fires. Only if path 2 is blocked can instances reach the normally quiet detector 2 with no interference, showing the bomb is there.

*Table 2. Non-physical detection (**)*

| Path | Existence Probability | Observation | |
|---|---|---|---|
| | | **No Bomb** | **Bomb (path 2)** |
| *Detector 1 by path 1* | 25% | Detector 1 fires | Detector 1 fires |
| *Detector 1 by path 2* | 25% | Detector 1 fires | Blows bomb |
| *Detector 2 by path 1* | 25% | Detector 2 never fires as out of phase path instances cancel out | Detector 2 fires** |
| *Detector 2 by path 2* | 25% | | Blows bomb |

*Non-physical detection,* registering a bomb that no physical photon can touch, is impossible in an objectively real world, but in our world it is a proven effect. We can detect a bomb sensitive to a single photon, without setting it off. A *counterfactual event,* a detector that could have fired but didn't, on a path the photon didn't physically travel, alters the physical outcome. Another tell-tale flaw exposes the physical world for what it is, an output of a non-physical quantum world.

## Entanglement

Bell used Einstein's reductio ad absurdum thought experiment (Einstein, Podolsky, & Rosen, 1935) to devise *Bell's inequality,* the definitive test of quantum theory's predictions vs. those of an objectively real world. If a Caesium atom releases two photons in opposite directions, quantum theory finds them "entangled", i.e. each acts randomly but the combination does not. The pair's start spin of zero is maintained by quantum mechanics as they evolve as a single system. No matter how far apart they get, if one photon is measured spin up, the other must be spin down. Yet each spin is random, so if one is up, how does the other *immediately* know to be down? In quantum mechanics, entangled photons are a single combination state, even if light years apart. To Einstein, that measuring one photon's spin instantly defines another's spin anywhere in the universe was "*spooky action at a distance*".

Testing Bell's inequality was one of the most careful experiments ever done, as befits the ultimate test of the nature of our reality. Quantum theory was proved right again when measuring one entangled photon resulted in another having the opposite spin, even though it was too far away for speed of light signaling (Aspect, Grangier, & Roger, 1982). That the combination state was maintained was proved beyond doubt, but no physical basis for this is possible:

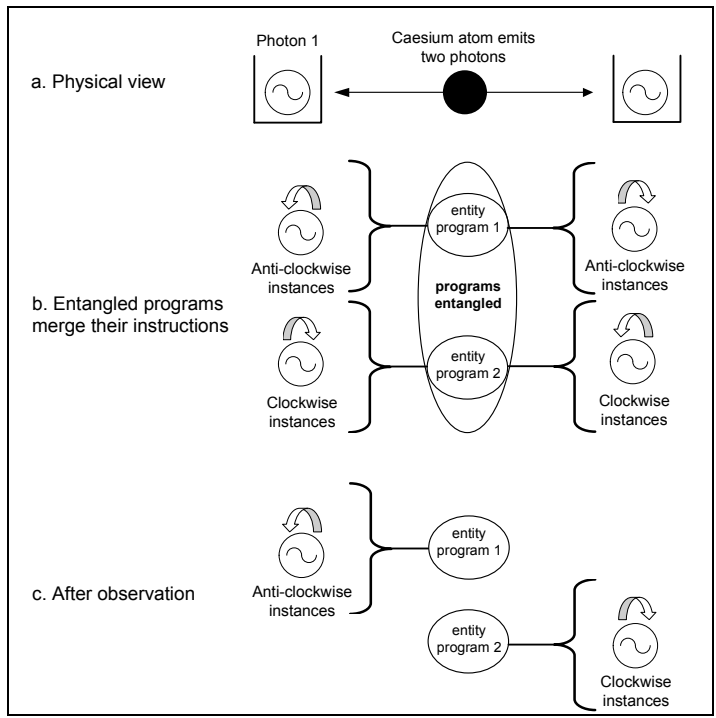

*Figure 24. Entanglement as program co-processing*

"*In short, the experimental verdict is in: the weirdness of the quantum world is real, whether we like it or not.*" (Tegmark & J. A. Wheeler, 2001) p4

Entangled states are now common in physics (Salart, Baas, Branciard, Gisin, & H., 2008) but make little sense in a purely physical reality. Two photons traveling in opposite directions are physically separate entities, so if each has random spin, as quantum theory says, what stops both being spin up (or down)? What connects them if not physicality? Why doesn't nature conserve physical spin by making one photon spin up and the other down? Apparently this is too much trouble, so it lets both photons have either spin, then as one is defined instantly adjusts the other to be the opposite, *regardless of where it is in the universe*.

Let quantum entities entangle as their programs merge after a reboot. In the case above, the initial merge state is two photons spinning opposite ways, i.e. one clockwise and one anti-clockwise. The merging that produced a processing overload then spreads out on the grid until another overload occurs. We see two photons going in opposite directions (Figure 24a), but in quantum terms it is one event served by the same code (Figure 24b). Both photons get clockwise and anti-clockwise instructions, so either may be observed with either spin. If a spin up instance restarts its entity program, it becomes a physically spin up photon and the entanglement stops, leaving the other photon with the opposite spin (Figure 24c). Initially, entangled instructions go equally to all instances, but if one program restarts in another node the remaining instances are of the opposite spin. The net spin remains zero because the same instructions exist at the end as at the start.

In entanglement, entity programs merge to jointly service a combination output that lasts until either program restarts in a new grid collision. It is non-local for the same reason that quantum collapse is, i.e. that program-to-node effects ignore node-to-node limits. No matter how far apart two entangled photons travel, they are connected not physically but at their merged program source. This can occur for any number of entity programs, e.g. Bose-Einstein condensates.

## The holographic interface

The holographic principle is that:

*Everything physically knowable about a volume of space can be encoded on a surface surrounding it* (Bekenstein, 2003).

For example, we can deduce depth because light travelling different distances arrives slightly out of phase. Flat photos just store light intensity but holograms also store the phase differences that encode depth, e.g. a credit card hologram of 3D image. This is done by splitting laser light, and letting the half that shines on the object interfere with a matched reference half to create an interference pattern (Figure 25). Light later shone on that pattern recreates the original 3D image.

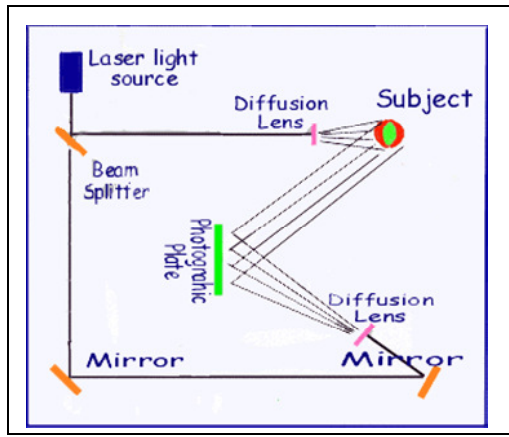

*Figure 25. Producing a hologram*[31]

We suppose that the information contained in a space depends on its volume, e.g. the number of memory chips in a space depends on its volume. Yet as they get smaller and smaller, to give more information, they eventually form a black hole, whose information, or entropy, depends on surface area not volume. So the holographic principle is maintained by the behavior of black holes (Bekenstein, 2003). While the universe indeed has three dimensions of movement, all the information we get about *any* physical object can be encoded on a two dimensional surface.

The holographic principle arises from this model as follows. If physical reality is an information transfer, there must be a transfer direction for the information to flow across and a dimension to express the information values transferred. As the grid proposed has four dimensions, two remain for the information to travel across. The virtual reality conjecture *requires* the holographic principle because information transfer in three-dimension must be across a two-dimensional surface. By the last chapter, each grid node can only get input from its nodal sphere of neighbors. The physical world registered at a point can be painted on the surface of the sphere around it because that is the structure that delivers it.

The holographic principle doesn't make the world two-dimensional. It is how we get information from the world, not how it is in itself. The screen generating the physical world has two dimensions but the world "out there" can have more, e.g. light moves with three degrees of freedom. That the world *presents* in two dimensions doesn't limit how it *operates*, any more than our 2D retina limits the world we see. Yet the physical world as an information hologram is not one we can walk around in like a detached observer, as our bodies *are* the holographic images. This is no Star Trek holo-deck to exit at will, as if we left where could we go? To leave this illusion would be to leave our physical existence.

**Heisenberg's uncertainty principle**

Heisenberg's uncertainty principle is that one can know a quantum particle's position or momentum exactly, but not both at once. Yet how can position and momentum be knowable, but both together unknowable? This *complementarity* is central to quantum mechanics, as choosing one measurement randomizes a following complementary one. In this model, a program entity "knows" or "observes" another by how it interacts with it:

"… *a measuring instrument is nothing else but a special system whose state contains information about the "object of measurement" after interacting with it:*" [23] p212

If all entities are waves, all interactions are essentially wave interference states. As sine waves follow De Broglie's inverse relation between momentum and wavelength[32], the uncertainty principle becomes that for a wave, one can know position or wavelength exactly, but not both at once.

---

[31] From http://www.mikecrowson.co.uk/Touching.html

[32] If p is momentum, $\lambda$ is wavelength and h is Planck's constant, then $p = h/\lambda$

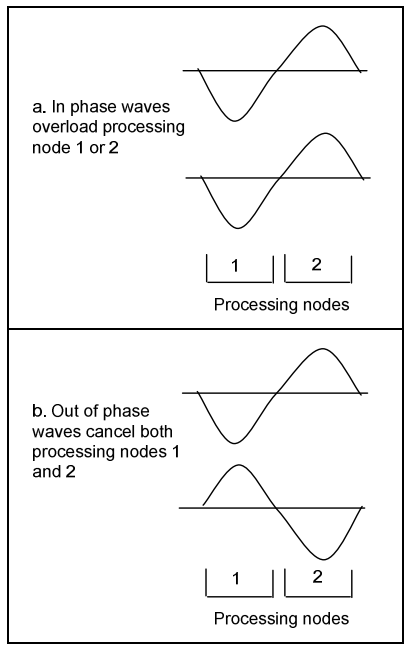

Figure 26 is a simple case of two-node waves interacting. If they are *in phase,* an overload at 1 or 2 gives position exactly but no wavelength information. Conversely if *out of phase,* they cancel entirely giving wavelength information but no node position. In both cases, the measurement event can't repeat because the waves have changed. A wave "observing" another gives position or wavelength but not both, with no repeats. If one measures position there is no wavelength data and if one measures wavelength there is no position data. In both cases, the measured wave has given all the information it has to the interaction.

The uncertainty principle, that position plus momentum information must exceed a minimum value defined by Planck's constant [33], is equivalent to saying that no quantum processing overload can be less than one Planck program.

### REDEFINING REALITY

This model questions the traditional reality of the physical world.

Figure 26. Waves interacting

### The measurement paradox

Science knows the world by observation but it collapses the wave function. How can science study quantum waves that by definition are unobservable, now and forever? The measurement paradox of quantum mechanics denies us in principle the opportunity to directly see the quantum wave, as any attempt to do so collapses it to an entity in one place. The information firewall of physical reality censors what we know and can know:

"*The full quantum wave function of an electron itself is not directly observable…*" (Lederman & Hill, 2004) p240

To this day, this issue is unresolved:

"*The history of the quantum measurement paradox is fascinating. There is still no general agreement on the matter even after eighty years of heated debate.*" (Laughlin, 2005) p49.

Many physicists accept logical positivism's statement that only "*…what impinges on us directly is real.*" (Mermin, 2009) p9. Yet if so, inherently unobservable quantum states can't be real, so are just convenient fictions, i.e. quantum theory references the unobserved so isn't scientific. Conversely, if quantum theory is scientific, then logical positivism is not a necessary condition for science.

Here, logical positivism is a nineteenth century reality myth masquerading as an axiom of science. Its tale of objects that inherently exist can't withstand scrutiny. An inherently existing object needs distinguishable left and right aspects to contain its self-existing extent. If so, it must be divisible into left and right parts. So a photon as a mini-object must have smaller parts, and these parts need still finer parts, and so on. If every object has smaller parts, how can it ever end? That physical objects always arise from others is like the earth sitting on the back of a giant turtle. As that turtle would need another turtle to stand on, ad infinitum, so every object would need sub-objects to comprise it. The universe can no more be "objects all the way down" than it can be "turtles all the way down"[34].

---

[33] Mathematically δx.δp ≥ ℏ/2 where x is position and p is momentum. ℏ is Plank's constant in radians, or Plank's constant divided by 2π.

The existential buck must stop somewhere and in this model it is light. Physical molecules split into physical atoms, but photons have no physical parts because a Planck program has no sub-programs. Quantum theory describes the camera that takes the pictures of physical reality. It can no more appear in its own photos than a finger can point to itself. To accommodate quantum mechanics, science should trade up to *research positivism*, that it should *predict* observables, from *naive positivism,* that it be *constituted* of them. Quantum mechanics is science, even though what it describes isn't physical.

## Many worlds theory

That quantum collapse is random means that no prior world events cause it. In 1957 Everett met this threat of an *uncaused cause* with many worlds theory, that every quantum choice spawns a new universe. If every quantum option actually occurs in an alternate universe, the multi-verse makes no choices and so stays deterministic. Everett invented his multi-verse machine just to contain the ghost of quantum randomness. While initially ignored, today physicists prefer it three to one over the Copenhagen non-view (Tegmark & J. A. Wheeler, 2001) p6. Yet billions of galaxies of photons, electrons and quarks each making billions of choices a second for billions of years means the:

"… *universe of universes would be piling up at rates that transcend all concepts of infinitude.*" (Walker, 2000) p107.

So the time you took to read that quote would create untold billions of universes like ours. Many worlds theory offends Occam's razor by multiplying universes unnecessarily. The clockwork multi-verse is a reincarnation of the clockwork universe that quantum theory demolished a century ago. Deutsch's attempt to rescue this zombie theory[35] by letting a finite number of universes "repartition" after each choice just recovers the original problem, as what chooses which universes are dropped? Why indeed should the universe, like a doting parent with a video-camera, copy everything we *might* do? The ex post facto argument of many worlds illustrates the ridiculous lengths positivists will go to explain away quantum theory. In contrast, in this model, choice is necessary to have information.

## The quantum paradox

A review of ten "myths" of quantum mechanics traces them all back to one core problem:

"*Thus, I conclude that the main reason for the existence of myths in QM {quantum mechanics} is the fact that QM does not give a clear answer to the question of what, if anything, objective reality is.*" (Nikolić, 2008) p43

Traditional objective reality began with Aristotle's idea that:

"… *the world consists of a multitude of single things (substances), each of them characterized by intrinsic properties …*" (Audretsch, 2004) p274

This two thousand year old view of a world of "things" with intrinsic physical properties existing in locations that limit their effects still dominates thought today. Officially quantum mechanics doesn't challenge it, but unofficially its immaterial waves spread and disappear with little regard for spatial locality. Yet if quantum mechanics is always right:

---

[34] In the apocryphal story, a scientist lecturing that the universe depended on nothing else was challenged by a little old lady, who said it sat on the back of a giant turtle. He laughed, and asked her what the turtle was standing on, but got the reply "*Sonny, it's turtles all the way down*".

[35] Zombie theories make no new predictions and can't be falsified. Like zombies, they have no progeny nor can they be (theoretically) "killed".

"*… why not simply accept the reality of the wave function?* (Zeh, 2004) p8

It isn't that easy, as if the wave function is real, then as Penrose says:

"*Thus, if we are to take $\psi$ as providing a picture of reality, then we must take these jumps as physically real occurrences too…*" (Penrose, 1994) p331

Yet we can't eat the quantum cake and have the physical world too. Schrödinger tried to treat wave states as a physical property, like say an electron's charge, but failed, and so has every else who has tried since then. The fact is that what quantum theory describes is nothing like the physical world we know. Quantum states disappear at will, so don't have the permanence of physical matter. Entangled effects ignore speed of light limits, so don't follow the laws of physical movement. Superposed states simultaneously exist in physically contradictory ways, so don't clash like matter. In sum, the *quantum world is in every way non-physical*. The quantum wave of an unobserved electron can spread across a galaxy then instantly collapse to a point, so as Barbour says:

"*How can something real disappear instantaneously?*" (Barbour, 1999) p200

When Pauli and Born took the quantum wave to be a probability of existence amplitude, physics ceased to be about anything physical at all:

"*For the first time in physics, we have an equation that allows us to describe the behavior of objects in the universe with astounding accuracy, but for which one of the mathematical objects of the theory, the quantum field $\psi$[36], apparently does not correspond to any known physical quantity.*" (Oerter, 2006) p89

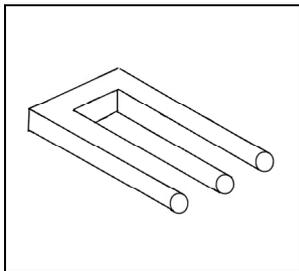

If quantum states are unreal because they are non-physical:

"*Can something that affects real events … itself be unreal?*" (Zeh, 2004) p4.

Or as Penrose says:

"*How, indeed, can real objects be constituted from unreal components?*" (Penrose, 1994) p313

*Figure 27. A paradox*

The quantum paradox is that quantum unreality creates physical reality. For nearly a century, physics faced it like a deer in headlights, paralyzed by the Copenhagen view that something is and isn't there. Yet paradoxes merely imply incompatible assumptions, e.g. Figure 27 has two square prongs and three circular ones, which is impossible. The paradox arises if a line can bound a square and circular prong at once, which it can't. Equally impossible is that quantum and physical realities co-exist, as the Copenhagen doctrine claims. This is as impossible as a line bounding two objects at once. Either the quantum world has a physical derivation or the physical world is a quantum derivative. As the former doesn't work, the latter it must be, i.e. the physical world is a quantum output. If so, is that output the primary reality?

## Non-physical realism

Bell's experiment assumed the following explicit world axioms (D'Espagnat, 1979):

1. *Physical realism.* That "*there is some physical reality whose existence is independent of human observers.*" (D'Espagnat, 1979) p158

2. *Einstein locality.* That no influence of any kind can travel faster than the speed of light.

3. *Logical induction.* That induction is a valid mode of reasoning.

---

[36] $\psi$ is the quantum wave function.

By the results, one or more of the above *must be wrong.* If realism and induction are true, then locality must be wrong. If locality and induction are true, then a real world can't exist independent of our observation of it. Physics has still not resolved this issue:

"*According to quantum theory, quantum correlations violating Bell's inequalities merely happen, somehow from outside space-time, in the sense that there is no story in space-time that can describe their occurrence*:" (Salart et al., 2008) p1

The resolution proposed here is to move the word "physical" from the realism definition to the locality definition. Realism then becomes:

> *that there is a ~~physical~~ reality whose existence is independent of human observers*

and locality becomes:

> *that no <u>physical</u> influence of any kind can propagate faster than the speed of light.*

This drops universal locality but keeps physical locality, i.e. limits Einstein's logic to physical objects. It also drops physical realism but keeps realism, i.e. permits a non-physical quantum reality. For example, a definition of realism like this:

"*If one adopts a realistic view of science, then one holds that there is a true and unique structure to the physical universe which scientists discover rather than invent.*" (Barrow, 2007) p124

Becomes instead this:

"*If one adopts a realistic view of science, then one holds that there is a true and unique structure to the universe which scientists discover rather than invent.*"

Science discovered rather than invented quantum states, so they are real, despite naive positivism. There really is world apart from us, but it is an unseen quantum world not the physical world we see. As the German philosopher Kant so acutely observed, we don't see things as they are in themselves (Kant, 2002) p392. We see the *phenomena* of physical reality not the *nuomena* of quantum waves. Shifting the locus of reality from the physical to the quantum world accommodates Bell's findings.

### Observation creates physical reality

A corollary of the above is that observation creates physical reality, as quantum physics has been telling us for some time now. The physical world isn't an objective reality we see impartially from above, like a bird. We see as a frog does, from the ground, as a participant. In system terms, we are "embedded observers", unable to see relativistic changes of time or space because they change us too.

Symmetric interactions don't differentiate observer and observed, so if we observe a photon it also "observes" us. Any interactions collapse the quantum wave, not just ours. Observation is unique in quantum theory because it creates the physical world, not because we are special observers.

In an objective world, that exists in and of itself, for observation to create reality makes no sense. In such a world, quantum theory could not be. Yet in our world, physical events like photon detection arise when observation collapses the quantum wave. We query the quantum database and a view of physical reality is thrown up on demand. Quantum theory tells us that the physical world we see as the ultimate cause is really an effect, as astronomy tells us that the earth circles the sun.

### The unseen world

A straw man positivists like to scapegoat is reifying quantum states, or making them physically real, but this model's *de-reification of physicality* isn't so easily dismissed. If the quantum world is real, physicality is no longer the reality touchstone. Expecting quantum states to be physical is like expecting a TV actor to have their onscreen persona in real life. It confuses cause and effect. The usual response to virtualism is to demand proof, and rightly so, but what exactly is the case that *all reality is the reality* *we see*? If one seeks proof, or even a reason beyond self-evidence, none is given. Remove the objective reality assumption, and positivism falls like the logical house of cards it is. If the physical world as a self-existent reality is a meta-physical opinion held, with no proof and despite contrary evidence[37], why is it unchallenged? Is it our physical bias?

"*Observers have to be made of matter…Our description of nature is thus severely biased: we describe it from the standpoint of matter.*" (Schiller, 2009) p834

The light we see is less than 1% of the electromagnetic spectrum. To see ultra-violet as bees do would show a different world. Likewise tuning a radio picks up radio waves beyond our senses. If instruments find our senses incomplete, why are instruments complete? Why hold so tightly to:

"… *the dogma that the concept of reality must be confined to objects in space and time…*" (Zeh, 2004) p18

By the logic of quantum theory, before our observational reality there is a quantum unreality of which the Copenhagen doctrine says *we must not speak*. Yet since entities only interact for an instant, they are in-between measurements more than in them:

"*Little has been said about the character of the unmeasured state. Since most of reality most of the time dwells in this unmeasured condition …the lack of such a description leaves the majority of the universe … shrouded in mystery.*" (Herbert, 1985) p194

If the world exists mostly in unobserved, uncollapsed quantum states, by what logic are only its brief moments of collapse real? *Surely reality is what is there most of the time?*

Or if quantum waves predict and cause physical reality, isn't making a cause "unreal" but its effect "real" backwards logic? If quantum states create physical states, by what logic are they unreal? *Surely reality is that which causes, not that which is caused?*

The current denial of quantum reality is doctrinal not logical, sustained by a blind faith in tenets of positivism, despite the evidence of non-physical quantum states and non-physical quantum collapse.

## Conclusion

By the findings of physics, the physical world *can't* an objective reality, but it *could* be a virtual one. This realization has been coming for some time. When matter was first attributed to unseen atoms, scientists like Mach didn't believe it. Then atoms were found to contain even smaller electrons, protons and neutrons, again unseen, and now science even recognizes unseeable quarks. Yet when quantum theory finds the ultimate base of reality is a probability, we cry "*Enough!*" This it seems is a step too far. How can the answer to life, the universe and everything be a number[38]? Yet now, after a thousand years of scientific progress, do we pull back and call it fiction?

It is to this place, that others shun, the virtual reality conjecture takes us, not to shock or amuse but to advance. It asserts what quantum theory implies, that a photon is an unseen probability of existence cloud that can instantly collapse over any distance, that physically arrives only when it is observed, that it defines its physical path after it arrives, and that it can be affected by choices that didn't happen. These cracks in the façade of physicality are real.

---

[37] That *we* only see the physical doesn't prove that *everything* is physical. Conversely, that there was a big bang proves that the physical universe is *not* a closed system, as most physicists believe it is.

[38] In Douglas Adams's Hitchhiker's Guide to the Galaxy, the computer Deep Thought, after a millennia of calculations, found that the answer to life, the universe and everything was 42. It was, of course, a joke.

We see ourselves in the sunlight of rationality standing before a dark cave of quantum paradox, but as in Plato's cave analogy, it is the other way around: we sit in a small cave with our backs to the quantum sunlight taking shadows cast on the wall of space as real. Quantum theory and relativity have loosed the chains of the objective reality illusion, but who can look away? Einstein did, but then waited for the quantum brilliance to fade, which it never did. Bohr walked out in his impenetrable Copenhagen suit, and saw only his own reflection. Since then, quantum theory has been in semi-quarantine, surrounded by a wall of arcane formulae, with those trained to enter brainwashed that *nothing here means anything* and that *everything here is imaginary*. Those who harness the new quantum light keep the old reality view intact by mathematical tools and semantic dogma, but it is a struggle.

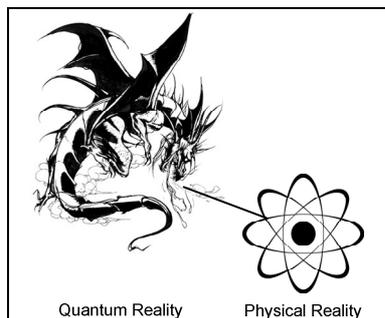

*Figure 28. The quantum smoke*

Yet quantum theory now makes no more sense than when first proposed. Let the new mantra be that *everything science describes is real.* Table 3 summarizes how quantum processing could create the physical reality of light. If quantum mechanics is Wheeler's *great smoky dragon*, then the physical world is just its smoke (Figure 28) (John A. Wheeler, 1983). The quantum world is no shadow world to the physical, but the world that creates what we see as a shadow.

*Table 3. Physical and processing properties of light*

| Physical property | Processing property |
|---|---|
| *Light.* A photon electro-magnetic wave: | *Processing.* A photon processing wave: |
| a) Sets absolute positive and negative values in space | a) Displaces the "surface" of our space absolutely |
| b) Is a sine wave that turns in "imaginary" space but moves in "real" space | b) Is a moving transverse rotation in quantum space that projects a sine wave perpendicular to our space |
| c) Moves at the fastest speed in any medium | c) Moves at the maximum cycle rate of the grid |
| d) Never fades in amplitude | d) Is maintained by ongoing grid processing |
| e) Conveys all its energy entirely at any wavelength point | e) Can deliver all its processing at any grid node |
| *Energy.* The energy a photon can deliver: | *Processing rate.* A Planck program's processing rate: |
| a) Decreases as its wavelength increases | a) Decreases as the program is shared by more nodes |
| b) Increases as its frequency increases | b) Increases as each node carries out the program faster |
| c) Must be an integer multiple of Plank's constant | c) Must be an integer divisor of one Planck program |
| d) Defines both Plank's constant and the size of space | d) Defines the size of both transverse and planar circles |
| *Quantum waves.* A quantum wave function can: | *Quantum instances.* Instances of an entity program can: |
| a) Spread outwards as a spherical wave | a) Distribute outwards as a spherical wave |
| b) Pass through two slits then interfere on exit | b) Pass through two slits then interfere on exit |
| c) Immediately "collapse" regardless of distance | c) Immediately disappear if the program restarts |
| d) Become a physical event with probability that depends on the net power of the wave at each point | d) Cause an entity program to restart, depending on the net processing instructions run at the node |
| *Quantum spin.* A photon polarized in one plane "exists" in other polarization planes, according to angle | *Quantum spin.* A photon's orthogonal quantum extent changes as it spins on its axis, according to angle |
| *The law of least action.* A photon detected at a point always takes the path of least action to that point | *The law of all action.* Photon instances take every path to a detector, and the first reboot becomes the physical photon |
| *Retrospective action.* A photon decides the path it took to a detector after it arrives | *Just in time action.* Photon program instances take every path so can restart the entire photon at any detector |
| *Non-physical detection.* Lack of interference can prove | *Quantum detection.* A detector blocking an alternate path |

| there is an obstacle on a path not physically traveled | blocks the instances that produce interference |
|---|---|
| *Superposition.* A quantum entity can exist in a combination of states that are physically incompatible | *Superposition.* An entity program can divide its processing existence among physically incompatible instances |
| *The measurement problem.* Our observation of the world creates the physical reality we get | *The measurement problem.* The query we call observation creates the information transfer we call physical reality |
| *Entanglement.* For entangled photons, one outcome affects the other, anywhere in the universe, and instantly | *Programs merge.* Entity programs merge to share instructions but reboot separately, regardless of grid node |
| *Holographic principle.* All the information a volume of space delivers can be encoded on its surface | *Transmission principle.* All the information a 3D network node receives must come from a 2D neighbor surface |

## QUESTIONS

The following questions highlight some of the issues raised:

1. Could electro-magnetic waves, including visible light, oscillate in a physical direction?
2. What does the entire electro-magnetic spectrum have in common?
3. Does the "imaginary" dimension of complex numbers actually exist?
4. Why does light uninterrupted never fade?
5. Why is the speed of light a maximum for any medium?
6. What is energy in processing terms?
7. Why does all energy come in Planck units?
8. How can a light wave deliver all its energy instantly at a point?
9. How can one photon go through both Young's slits at once?
10. How can a quantum wave collapse instantly to a point, regardless of its spatial extent?
11. What are counterfactuals? Do they exist?
12. Is a photon a wave or a particle or both?
13. How can a photon of polarized light pass <u>entirely</u> though a filter nearly at right angles to it?
14. What is the law of least action and how is it possible?
15. Why is a photon's spin on any axis always the same Planck values?
16. Is non-physical knowing, or knowing a thing without physical contact, possible?
17. How can a photon choose the physical path it took to a detector *when* it arrives?
18. Why don't physically incompatible quantum states clash?
19. Will we ever be able to see quantum waves directly?
20. If the physical world is a virtual reality, why then must the holographic principle be true?
21. How can entangled photons instantly affect each other anywhere in the universe?
22. Do we create physical reality by observation? If so, is the world just a dream?
23. If quantum states produce physical states, which are real? Can both be real?
24. Where do random quantum choices come from?

## ACKNOWLEDGEMENTS

I thank Alex Whitworth, Matthew Raspanti and Alasdair Broun for insightful comments.